\documentclass{emulateapj}
\usepackage{longtable}
\usepackage{graphicx}

\newcommand{\ha}{H$\alpha$}

\newcommand{\smy}{$\rm {\it h}_{100}^{-2}~ M_\odot ~ yr^{-1}$}
\newcommand{\ewmin}{10~\AA}

\newcommand{\h}{$h^{-1}_{100}$}

\newcommand{\cj}{CL~J0023$+$0423B}

\slugcomment{Submitted to ApJ}
\shorttitle{\ha \ SFRs of \cj}
\shortauthors{Finn, Zaritsky, \& McCarthy}

\begin{document}
\title{H$\alpha$-Derived Star-Formation Rates For The $z = 0.84$ 
Galaxy Cluster CLJ0023+0423B\altaffilmark{1,2}}

\author{Rose A. Finn\altaffilmark{3,4}, Dennis Zaritsky, and 
Donald W. McCarthy, Jr.}
\affil{Steward Observatory, 933 N. Cherry Ave., University of Arizona, Tucson, AZ  85721}
\email{rfinn@astro.umass.edu; dzaritsky, mccarthy@as.arizona.edu}
\altaffiltext{1}{Based on observations with the MMT Observatory,
a joint venture of the Smithsonian Astrophysical Observatory and
the University of Arizona.}
\altaffiltext{2}{Based on observations with the NASA/ESA {\it Hubble Space
Telescope}, obtained at the Space Telescope Science Institute, which is
operated by AURA, Inc.}
\altaffiltext{3}{NSF Astronomy and Astrophysics
Postdoctoral Fellow}
\altaffiltext{4}{Current address:  Department of Astronomy, University of Massachusetts, Amherst, MA  01003}

\begin{abstract}
We present \ha-derived star-formation rates (SFRs) for the galaxy cluster 
CL~J0023+0423B at $z = 0.845$.  Our $3\sigma$ flux limits corresponds 
to a star-formation rate of 0.24~\smy, and our minimum reliable 
\ha~+ [N~II] equivalent width is $> 10$~\AA, demonstrating  
that near-infrared narrow-band imaging
can sample the star-forming 
galaxy population in distant clusters.
Comparison with spectroscopy shows that the number of false detections
is low ($9 \pm 6$\%) and that our \ha \ equivalent widths 
are correlated with 
spectroscopically determined [O II] equivalent widths.
A magnitude-limited spectroscopic survey conducted over the same
area missed 70\% of the star-forming galaxies and
65\% of the integrated star formation.
Using Hubble Space Telescope Wide Field Planetary Camera 2 Archive images, 
we fit Sersic profiles to all galaxies with significant
narrow-band equivalent widths and find that 
equivalent width decreases
as the steepness of galaxy profile increases. 
We find no significant population of early type galaxies
with ongoing star formation.
The integrated SFR per cluster mass of \cj \ is 
a factor of ten higher than that of the three $z \sim 0.2$ clusters 
in the literature with available \ha\ observations.  
A larger sample of $z \sim 0.8$ clusters spanning a range of cluster masses
is needed to determine whether this variation is 
due to a difference in cluster mass or redshift.
\end{abstract}


\section{INTRODUCTION}

The observed variation of galaxy properties as a function of environment
\citep{hubble31} is evidently an important clue to how galaxies form and evolve.
The two classic observations that drive this field of study are that 
nearby galaxy clusters contain a higher fraction of E/S0 galaxies 
than observed in the field
\citep[e.g.][]{dressler80,whitmore93} and that the fraction of blue galaxies
in a cluster increases as a function of redshift \citep{butcher84}. 
The subsequent 
interpretation of these results has been varied, often conflicting, and
not yet distilled. The difficulty, in part, lies in the lack of direct 
observations that trace star 
formation in large samples of cluster and field galaxies.

Broad-band colors are most often used to trace star-formation rates 
\citep[for example,][]{butcher84}, 
but these offer a crude measure and are affected by dust and metallicity. 
Comparisons of emission-line measurements of star formation are complicated
by the use of different lines at different redshifts, by the difficulty
in getting large samples at high ($> 0.5$) redshifts, and by the dust and 
metallicity
sensitivity of [O~II], the most popular indicator at high redshift.
We present our initial attempt to compile a sample of \ha\ 
measurements of star formation in a sample of high-redshift 
($\sim 0.7 - 0.8$) galaxy clusters by presenting the results for 
our first completed cluster, \cj.  The advantages of these 
\ha\ data are that they are directly comparable to $z < 0.3$ studies, 
they directly measure the ionizing flux from young stars, and 
they are less sensitive to extinction than the common [O~II] measurements.

Even at low redshift, \ha\ observations of cluster populations are limited.
\citet{kenn83b} and \citet{kenn84} use 
\ha \ spectroscopy to study 
the star-formation properties of four nearby clusters and 
find that cluster and field spirals of a given
morphology have similar star-formation rates (SFRs) in three of four clusters.
In contrast, an objective prism study of nearby Abell clusters, 
\citet{moss93,moss00} 
find that the difference in star-formation properties between field and 
cluster spirals depends on morphology.
Cluster \ha \ studies have now been pushed out to redshifts of $0.1 - 0.3$ using
imaging \citep{balogh00} and spectroscopy \citep{couch01,balogh02}.

At intermediate and high redshift the results have depended on the less robust
[O II] measure of star formation. The CNOC \citep{balogh97,balogh98} and 
MORPHS \citep{smail97,dressler99, poggianti99} surveys agree that
cluster galaxies of all Hubble types have lower SFRs than the
same type field galaxies. However, 
photometric modeling of the CNOC clusters favors a 
slow decline in star formation \citep{ellingson91, kodama01b} as one approaches
the cluster, while
the MORPHS spectroscopy reveal a large population
of starburst and post-starburst cluster galaxies, which reflect sudden and
dramatic changes in star formation rates. 
In a cluster at $z = 0.83$, \citet{vandokkum99} 
find that although the observed merger rate
is significantly higher than the field, there is no sign of 
excess star formation.  
\citet{postman98,postman01} study four $z \sim 0.9$ clusters
and find that cluster galaxies have systematically lower
SFRs than field galaxies at similar redshifts. The only result
that appears to be consistent throughout all of these studies and across 
redshift
is that cluster galaxies have lower rates of star formation than field 
galaxies.  The origin of this empirical result is still debated.

Given this situation, one might naturally conclude that significant progress
could be made at any redshift, and that all things being equal one
should focus at low redshift. However, there are two developments
at low redshift that will vastly improve the current situation. 
SDSS and 2dF have already begun to provide \ha\ measurements
of star formation in local galaxies as a function of environment \citep{gomez03, 
lewis02},
and an ongoing deep \ha\ selected survey of nearby clusters will complete the 
picture \citep{sakai02}.
For high-redshift, one wants to push as high as possible to maximize the
evolutionary effects.
The redshift regime of $0.7 - 0.8$ is the highest 
for which there exists a significant number of clusters with complementary
data, such as redshifts and morphologies;  however, \ha\ moves beyond the 
optical window into the $J$ band. 

To observe the \ha\ line, 
we are undertaking a near-infrared, narrow-band \ha \ imaging
survey of ten $z \sim 0.8$ clusters. 
Here we present results for the first
cluster in our sample, \cj.
The goals of this initial paper are to demonstrate the
accuracy and sensitivity of the technique and to 
provide the first comparison between \ha \ imaging results for
a cluster at $z \sim 0.8$ and similar surveys of
lower-redshift clusters.

\begin{figure}
\plotone{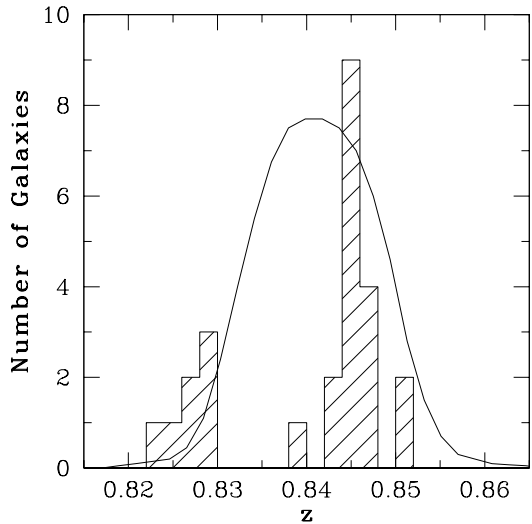}
\caption{Trace of narrow-band filter (solid line) plotted with 
shaded histogram of spectroscopic redshifts from \cite{postman98}.}
\label{jcont}
\end{figure}

We choose CLJ0023+0423B as part of our cluster sample 
because its redshifted \ha \ emission lies in a 
standard near-infrared narrow-band filter (Figure \ref{jcont})  
and because it has Hubble Space Telescope ($HST$) 
Wide Field Planetary Camera 2 (WFPC2) imaging and 
Keck spectroscopy \citep{postman98,lubin98a}.
These additional data provide a necessary complement to
our imaging. For example, 
\citet{postman98} present 
a mass estimate for the cluster, $\rm \sim 1-5\times 10^{14} ~{\it 
h}^{-1}_{100}~ M_\odot$,
and  \citet{lubin98a} use WFPC2 imaging to measure an 
early-type fraction of $\le 33\%$, a value
comparable to that of the field.  

The organization of this paper is as follows.  In
\S\ref{clj0023_reduction} we describe
the observations and data reduction procedure.
In \S\ref{clj0023_fluxcal} we explain the 
calibration of the narrow-band images, and
in \S\ref{clj0023_contsub} we describe 
the continuum-subtraction process and model 
the associated uncertainties.  Results from the narrow-band imaging
are presented in \S\ref{clj0023_results}, and we
discuss these results in the context of existing
observations in 
\S\ref{clj0023_discussion}.  The paper is
summarized in \S\ref{clj0023_summary}. 
We assume $\Omega_0 = 0.3$ and $\Omega_\Lambda = 0.7$ throughout 
and express results in terms of $h_{100}$, where 
$\rm H_0 = \it h_{100} \ \rm {100~km~s^{-1}~Mpc^{-1}}$.

\section{OBSERVATIONS \& DATA REDUCTION}
\label{clj0023_reduction}
We observed CL~J0023+0423B using the PISCES
near-infrared camera \citep{mccarthy01} at the 6.5m MMT on 2002 December 19
equipped with a standard $J$-continuum narrow-band filter 
($\lambda_c = 1.208$\micron, see Figure \ref{jcont}).
A $J$-band image, appropriately scaled (see \S \ref{clj0023_contsub}), is used for continuum
subtraction. Individual 
exposures are 90 seconds for the $J$-band and ten minutes for the narrow-band.
The total integration
time is 100 and 22.5 minutes for the narrow and $J$-bands respectively.
The telescope is dithered
between successive images in increments of 10 to 15\arcsec.  
Conditions were photometric and the seeing was 0.9 arcsec. 
PISCES does not yet have a guiding capability at the MMT,
so each of the ten minute exposures is unguided, which results in slight
degradation  of the image quality (an effective seeing of 1.1 arcsec).
The pixel scale of PISCES at the MMT is 0.18 arcsec/pixel with an inscribed
circular field of 3.1 arcmin in diameter which corresponds to 1.0~\h~Mpc at
$z = 0.845$.

The image processing follows a standard procedure.
PISCES images exhibit a cross-talk phenomenon where 
each pixel in one quadrant leaves a negative imprint and tail 
at the identical positions in the three other quadrants 
(see \citeauthor{mccarthy01} \citeyear{mccarthy01} for more details).
We first correct the images for this cross-talk  
and then subtract the dark frame.
Dividing by dome flats removes small scale pixel sensitivity variations.
To remove large scale gradients, we divide the object frames
by an image that is the result of median combining object frames
using an upper 
rejection threshold  of $\sim$100~ADU above the maximum sky level.
A bad pixel mask is used to eliminate these pixels from
being included in the construction of the final combined
frame.  We map the geometric distortion in the $J$ and narrow-band filters
using MMT/PISCES observations of the open cluster NGC~1193 taken
on 2002 October 13, 
the United States Naval Observatory star catalog \citep{monet03}, and
the IRAF task GEOMAP.  The residuals from a fifth order Legendre 
polynomial astrometric fit in both
the x and y directions have
an RMS of 0.5\arcsec \ along each axis.
We correct the images for this distortion using 
the IRAF GEOTRAN task. 
The object frames are aligned and combined, and the final
narrow and $J$-band images are aligned using GEOTRAN.
The combined $J$ and narrow-band images have an effective seeing FWHM of
1.1\arcsec, or six pixels.

The final narrow-band images require some slight further processing to ensure
uniformity in the background. While the 
combined $J$-band image shows peak-to-valley variations
that are $\le 0.05\%$ across the entire array,
the combined narrow-band image has
large-scale variations in sky level that are $\le 0.2$\%.  We use SExtractor 
\citep{bertin96} to create an image of the residual background variation.  
Unsure of the origin of these remaining variations in sky level, we 
try both subtracting and normalizing by the background image.
Re-flattening the data produces fewer spurious 
emission and absorption detections, so we use this technique to 
remove the residual, large-scale sky variations.
The resulting narrow-band image is flat
to within $\le 0.07$\% across the entire image.

Because the PISCES camera has an inscribed circular field, some corners
in the combined frames have not been fully illuminated in each exposure.
We exclude these underexposed, lower signal-to-noise areas from our 
analysis.  

\subsection{Source Detection and Photometry}
We use SExtractor for source detection and photometry with parameters set
to ensure detection of all objects that are visually detected in the combined $J$ 
$+$ narrow-band image.
The adopted selection parameters are 
a signal-to-noise threshold of 2.0$\sigma$ per pixel where $\sigma$ is the 
standard deviation in sky, a minimum object
area of 12 pixels,  a tophat 5x5 convolution kernel, and
a background mesh size of 48 pixels.
Source positions and apertures are determined
using the combined $J$ and narrow-band image
and then applied to the $J$, narrow, and narrow $-$ scaled $J$-band images
as described below.
We measure aperture magnitudes in each of these three images
using an aperture defined by the outer isophote of the detection threshold
in the $J$ + narrow image. 
These detection criteria correspond to a $J$-band total magnitude limit
of $22.5$ as determined by the SExtractor MAG-AUTO algorithm.

The \ha\ flux limit is set by the minimum object size and the noise
in the continuum-subtracted image.
The average standard deviation of the background calculated in
several $15\times 15$ pixel boxes is  
0.011~$\rm ADU~s^{-1}$. Therefore, the $1\sigma$ 
noise associated with a sum over the minimum 12 pixel
object area is 0.038~$\rm ADU~s^{-1}$,
and our adopted $3~\sigma$ detection threshold is 0.11~$\rm ADU~s^{-1}$.
This limit corresponds to the smallest possible object area of 12 pixels.  
The median object area is 54 pixels, and the associated $3\sigma$ error
is 0.24~$\rm ADU~s^{-1}$.

\subsection{Flux Calibration of $J$- and Narrow-band Images}
\label{clj0023_fluxcal}
Converting continuum-subtracted narrow-band fluxes to
SFRs requires knowing 1) the relative throughputs of the $J$
and narrow-band filters, 2) the calibration of the $J$-band flux to
$\rm ergs \ s^{-1} \ cm^{-2}$, and 
3) the conversion factor from photons~s$^{-1}$ to 
SFR for
\ha\ photons.  We relate the narrow-band filter to the 
$J$-band filter, which is part of a standard magnitude system,
by observing solar-type standard stars from  \cite{persson98}  through both
filters.
We use these data and the IRAF DAOPHOT package to solve the $J$-band 
photometric transformation
equation, solving only for the zeropoint and airmass terms (see 
Table \ref{phot} for a solution of 
PISCES calibration data taken on 2002 December 19).
Narrow and $J$-band images are corrected for airmass using
the $J$-band extinction.  This is appropriate because we 
find the ratio of narrow-to-$J$
fluxes to be independent of airmass for the \cj \ filter.
To calculate the expected ratio of efficiencies between the two filters
we integrate the Planck function for $T = 5800$~K over both 
the $J$ and narrow band filters, multiplying 
by the filter and 
atmospheric transmissions at each wavelength.
We assume that the detector quantum efficiency is constant 
over both filter bandpasses. 
The calculated ratio of narrow-to-$J$ photons is 0.0507
for a solar-type star.
We repeat the calculation using a model solar spectrum \citep{kurucz79} 
and obtain  consistent results.  We adopt the Planck function
because of its convenient functional form.

We convert the $J$-band flux of a standard star 
to Janskys using a $J$ magnitude zeropoint of 1600 Jy \citep{campins85},
then to ergs~s$^{-1}$~cm$^{-2}$ by multiplying by the $J$ filter 
bandwidth of 0.3~\micron,
and then to photons~s$^{-1}$~cm$^{-2}$ 
by dividing by $hc/\lambda_c$, where the 
central wavelength of our $J$ filter is 1.25~\micron.
The predicted photons~s$^{-1}$~cm$^{-2}$ in the narrow-band filter
is the product of the $J$-band flux times
the ratio of the narrow-to-$J$ integrals.

To convert to SFR, we first correct fluxes by 1 magnitude
for dust extinction \citep{kenn83} and scale by 0.77 to 
correct for [N~II] contamination \citep{tresse99}.
We convert to \ha \ luminosity assuming
all sources are at the cluster redshift, and we
relate \ha \ luminosity to SFR using
\begin{equation}
\rm
1~ ergs \ s^{-1} = 7.9 \times 10^{-42}~ M_\odot ~ yr^{-1} 
\end{equation}
\citep{kenn94}.  
Using a large sample of local galaxies drawn from the SDSS, 
\citet{brinchmann03} show that 
the Kennicutt SFR conversion is robust on average.
However, the conversion varies by a factor of 2.5 from the lowest to highest
mass galaxies.  We do not have the additional spectral information required to
fine-tune the SFR conversion, so our SFRs have an associated 
$\sim$30\% uncertainty that we add in quadrature to 
other sources of random error.

The average narrow-band flux zeropoint determined from three
standard stars is listed in Table \ref{phot} in units of 
$\rm ergs \ s^{-1} \ cm^{-2}$ and in SFR limits in units of \smy.
Our $3\sigma$ detection threshold of 0.11~$\rm ADU~s^{-1}$ corresponds to
0.24~\smy \ at $z = 0.845$.  
However, our detections are also limited in equivalent width,
as discussed in \S\ref{clj0023_contsub}.

\section{CONTINUUM SUBTRACTION}
\label{clj0023_contsub}

\ha \ emission and absorption sources are defined to be sources
that have nonzero residual (narrow-band $-$ continuum) flux.
The $J$-band flux, scaled by the ratio of narrow-to-$J$ filter
throughputs, is adopted as the continuum flux
within the narrow-band filter. The determination of the 
relative filter throughputs is described in \S\ref{clj0023_fluxcal}.

We compare two methods for calculating continuum-subtracted 
narrow-band fluxes.
In the first method, we use SExtractor to measure fluxes in the narrow
and $J$-band images separately, and then difference the narrow-band
and scaled $J$-band fluxes object by object.  
In the second method, we use SExtractor to construct a background
image for each filter, subtract these, and then subtract
the scaled $J$-band image from the narrow-band image.
We then run SExtractor on the continuum-subtracted image.
We scale the $J$-band image by 0.0488, which corresponds to the
ratio of narrow-to-$J$ throughputs assuming a flat spectral energy distribution.
The residual \ha\ fluxes 
from the two methods agree within $1\sigma$.  
We choose to work on the $J$ and narrow-band images separately
because negative sources (absorption) in the continuum-subtracted
image are not directly detected with SExtractor, and any modifications of
the relative throughputs of the filters is more difficult to implement.

\begin{figure}
\plotone{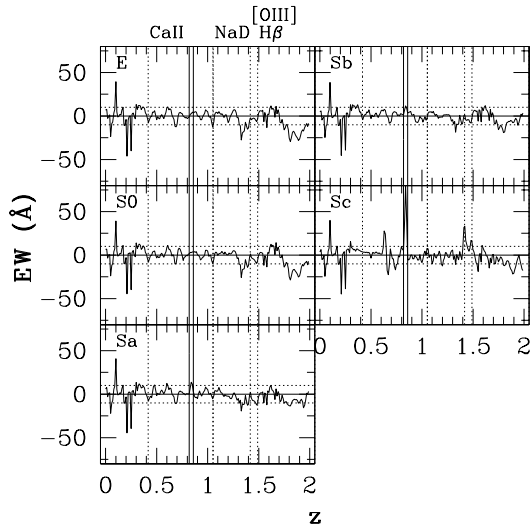}
\caption{EWs of E, S0, Sa, Sb, and
Sc galaxy types as a function of redshift.  Galaxy spectra are
from \citet{mannucci01}.  Solid vertical lines show redshift range where
narrow-band filter detects \ha.  Dotted vertical lines show redshifts
where other spectral features pass through filter window.  Low-redshift spikes
are numerical artifacts and do not correspond to real spectral features.}
\label{ewz}
\end{figure}

The scaled $J$-band flux provides an imperfect
estimate of the continuum within the narrow-band.
To quantify the error in our continuum estimates we calculate
the measured narrow-band equivalent width (EW) 
for five galaxy types, E through Sc, 
as a function of redshift using composite 
spectra from \citet{mannucci01}.  We define EW as
\begin{equation}
\label{eweqn1}
EW = \frac{f_n - r f_J}{f_J} \Delta\lambda_J,
\end{equation}
where $f_n$ is the narrow-band flux in $\rm ADU~s^{-1}$,
$r$ is the calculated ratio of narrow-to-$J$ filter
throughputs (0.0488),
$f_J$ is the $J$-band flux in $\rm ADU~s^{-1}$, and $\Delta\lambda_J$ is
the bandwidth of the $J$-band filter or 0.3~\micron.
With this definition, emission sources have positive EWs.
The EWs of the composite spectra are shown in 
Figure \ref{ewz} as a function of galaxy redshift.  
The standard deviations of the narrow-band EW for
the E, S0, Sa, Sb, and Sc
galaxies are 10.4, 9.8, 8.3, 8.3, and 12.3~\AA, respectively.
The standard deviation of the Sc galaxy includes the \ha \ 
emission line that we are trying to detect, 
so it is an overestimate 
of the contamination. 
We adopt 10~\AA \ as our minimum reliable EW and consider only
objects with EW $> 10$~\AA \ as significant detections of \ha\ flux.
The scatter in the measured
EWs is dominated by spectral features, not errors introduced by 
continuum slope across the $J$-band window.  Because, the 
\cj \ filter lies in the middle of the
$J$-band,  the continuum slope through the $J$-band 
does not severely undermine the calculated
narrow-band continuum.

\section{RESULTS}
\label{clj0023_results}
We discuss continuum-subtracted flux in terms of two quantities,
rest-frame equivalent width (EW$_R$) and SFR.    
We define EW$_R = $EW$/(1+z)$, where $z$ is the cluster redshift,
and we assume that all objects with significant EWs are associated
with the cluster.
The uncertainty in EW$_R$ is  given by
\begin{equation}
\sigma_{EW_R} = \frac{\Delta\lambda_J}{(1+z)} \sqrt{(\frac{1}{f_J})^2 \sigma_{f_n}^2 + (\frac{f_n}{f_J^2})^2 \sigma_{f_J}^2},
\end{equation}
where the narrow and $J$-band flux errors, $\sigma_{f_n}$ and $\sigma_{f_J}$, 
are the sum in quadrature of zeropoint
errors and SExtractor photometric errors.  

We calculate 
the SFR by scaling the continuum-subtracted
flux using the conversion from $\rm ADU~s^{-1}$ to \smy \ given in Table 
\ref{phot}.  Uncertainties are propagated and include the uncertainty
in the conversion to SFR.

We list EW$_R$ and SFR for every galaxy detected 
in the combined $J$ $+$ narrow-band
image in Table 2.
The columns are described in the Table notes.
The EW$_R$  is not corrected for
[N~II] contamination, but the flux, luminosity, and SFR are corrected.

\begin{figure}
\plotone{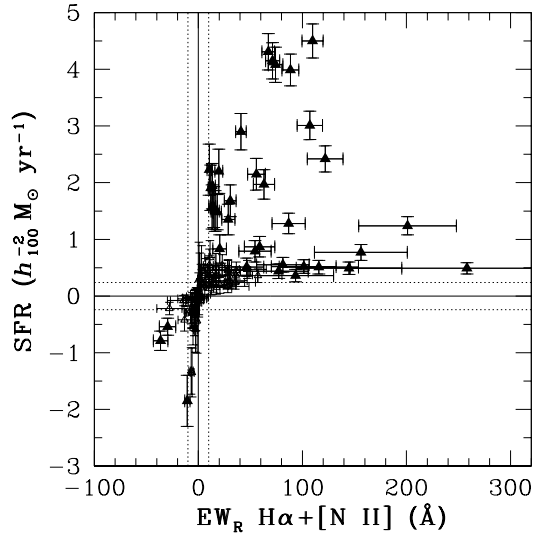}
\caption{SFR versus EW$_R$ for all galaxies in the CL~J0023+0423B field.
Dotted lines mark the $3\sigma$ SFR limits and 10~\AA \ EW$_R$ limits.
We consider all objects with $>3\sigma$ continuum-subtracted 
flux \ and $|EW_R| > 10$~\AA \
as significant detections.
}
\label{ewsfr}
\end{figure}

Our detection sensitivity is limited in two ways, first by the
noise properties of the final narrow and $J$-band images, 
and second by the uncertainty
in estimating the narrow-band continuum from the $J$-band flux.
We show the relationship between these two limitations in Figure \ref{ewsfr},
where we plot SFR versus EW$_R$.
We consider all objects with $>3\sigma$ continuum-subtracted 
flux \ and ${\rm |EW}_R| > 10$~\AA \
as significant detections.

\begin{figure}
\epsscale{1.}
\epsscale{.97}
\hspace*{-1.25cm}
\plotone{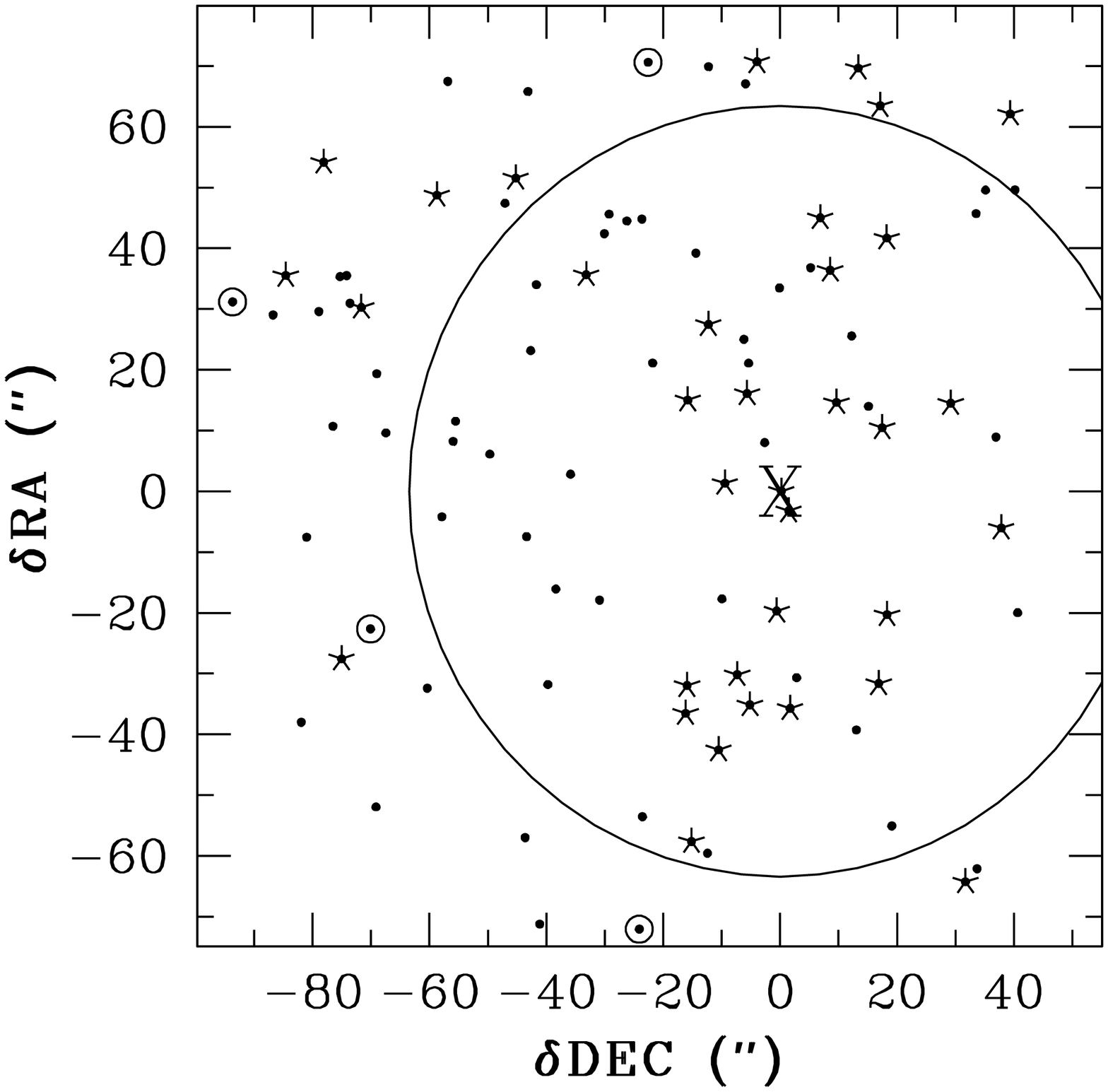}
\caption{(Top - see 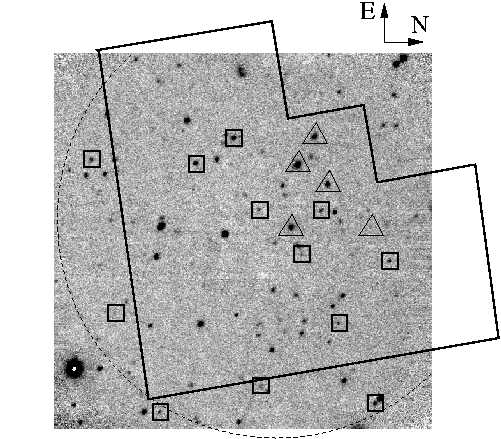) $J$-band image of CL~J0023+0423B.  Squares indicate
spectroscopically confirmed cluster members; triangles indicate 
spectroscopically confirmed members of $z = 0.827$ group \citep{postman98}.  
WCPC2 footprint is overlaid.  Data outside the dashed circle are excluded 
from analysis due to lower signal-to-noise.  (Bottom) Schematic
with same scale as top panel showing positions of galaxies 
relative to cluster center (X).  
Galaxy positions are marked with dots.
Galaxies with significant emission and absorption are marked with
stars and circles, respectively.  The circle 
shows $R_{200}$.}
\label{clj0023narrow}
\end{figure}

\begin{figure}
\epsscale{1}
\plotone{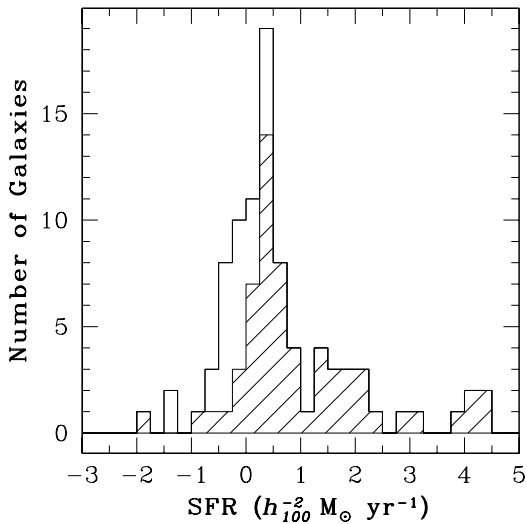}
\caption{Distribution of \ha-derived SFRs for galaxies
in the CL~J0023+0423B field.  Open histogram shows distribution of all 
galaxies, and the shaded histogram shows the
distribution for galaxies with $\rm |EW_R| >$~\ewmin.  
Negative SFRs indicate objects with 
significant narrow-band \ha \ absorption.}
\label{lfi}
\end{figure}

The spatial distribution of all galaxies and those with emission or absorption
is shown in Figure \ref{clj0023narrow}.
The top panel of the Figure is the final 
$J$-band image of \cj.  The bottom panel is a schematic of galaxy positions
relative to the cluster center.
We define the cluster center to be coincident with the brightest
galaxy and mark the position on the schematic with an X.
We detect a total of 94 galaxies, of which 35 have 
significant \ha \ emission and 4 have significant narrow-band 
absorption.
The distribution of SFRs is shown in Figure \ref{lfi}.  The integrated
SFR for all galaxies with EW$_R >10$~\AA \  and $> 3\sigma$ 
continuum-subtracted flux is 
68$\pm$22~\smy \ with an average 
SFR of 1.9~\smy \ for all galaxies with EW$_R >10$~\AA \  and $> 3\sigma$ 
continuum-subtracted flux.

\subsection{Cross-Check with Spectroscopy}

\cite{postman98} have published spectroscopy for 127 objects in
the field of CL~J0023+0423B.
Seventeen are confirmed cluster members at $z = 0.845$ 
and another 7 are associated with a group at $z = 0.827$.  
Figure \ref{jcont} shows the trace of
the narrow-band filter plotted with spectroscopically confirmed
cluster members from \cite{postman98}.
All spectroscopically confirmed members lie just redward
of peak transmission of the narrow-band filter, and the lower redshift
group lies in the blue wing of the filter.
However, not all of the spectroscopic members lie
within our \ha \ image.
Spectroscopic coverage of \citet{postman98} matches
our $J$-band image in the East-West direction and extends an additional 2.3 arcmin
to the South and additional 2.5 arcmin to the North.  
The squares in the top panel of Figure \ref{clj0023narrow} 
show the positions of the eleven spectroscopic
cluster members on our $J$-band image.  An additional spectroscopic member 
is detected but lies outside the area of full illumination (dashed circle).
Therefore we do not include this source in our analysis.

\begin{figure}
\plotone{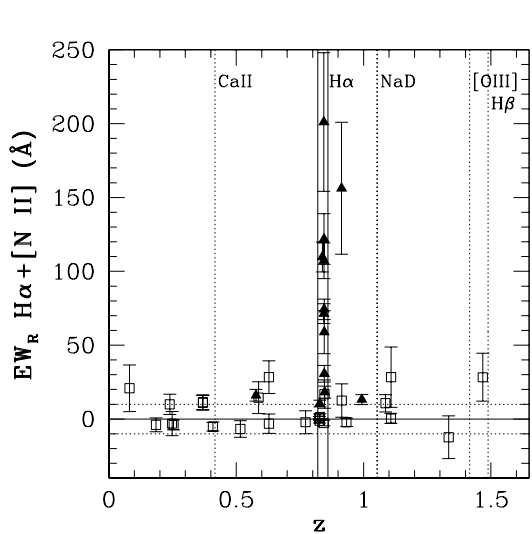}
\caption{EW versus redshift
for galaxies with spectroscopic redshifts from \cite{postman98}.
Galaxies with $\ge 3\sigma$ continuum-subtracted fluxes 
are depicted with solid triangles, and galaxies with $< 3\sigma$ 
continuum-subtracted fluxes with open squares.    
After applying the EW and flux cuts we find three false detections from
non-cluster galaxies.}
\label{haz}
\end{figure}

We have matched 39 spectroscopic targets with well-determined
redshifts to galaxies in our $J$-band image, eleven of which
are the clusters members described above.  
In Figure \ref{haz} we plot the 
EW$_R$ versus spectroscopic redshift for these galaxies.  The
bandpass of the filter is marked with solid lines, and dashed 
horizontal lines show the $|{\rm EW}_R| > 10$~\AA\  limit. 
Galaxies above our EW$_R$ and $3\sigma$ flux limits are shown
with solid triangles.  Galaxies below these limits
are shown with open squares for completeness, but we do
not consider these to be significant detections.
We find three significant ($\ge 3\sigma$) detections
from non-cluster galaxies.  In
Figure \ref{haz}, we overlay vertical dashed lines showing where prominent
spectral features pass through our narrow-band filter.  
None of the detections is attributable
to emission or absorption
from another spectral feature.  The galaxies may have non-standard
spectral energy distributions such as broad 
emission due to an active galactic nucleus.
From this comparison,
we conclude that contamination from non-cluster members
is 9$\pm$6\% (3/32).  

We also use the spectroscopic data to test our assertion that
the scaled $J$-band flux is an adequate estimate of the narrow-band continuum.
Differences among galaxy colors, which will introduce errors 
in the inferred continuum, 
might correlate with EW if we have misestimated the $J$-band continuum.
Although, we find a larger dispersion in EW toward bluer colors,
which is a consequence of the decreasing signal-to-noise
with increasing $J$ magnitude, we find no systematic
effects.

\begin{figure}
\plotone{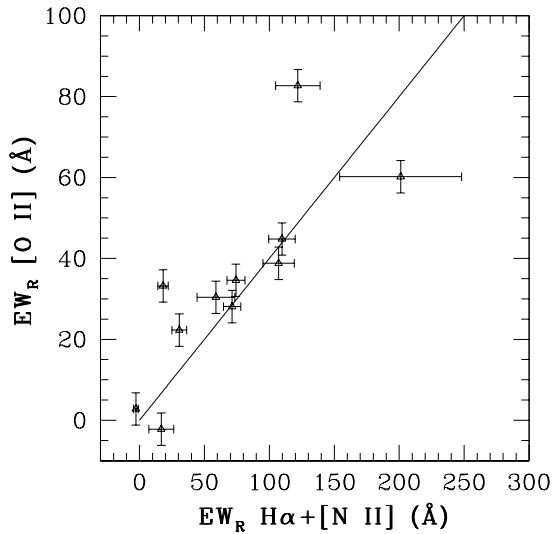}
\caption{Comparison between [O~II] and \ha\ EW$_R$
for eleven spectroscopically confirmed cluster members \citep{postman98}.  
Solid line shows empirical relation for local galaxies 
\citep{kenn92a,kenn92b}.}
\label{sfrhao2}
\end{figure}

Finally, we use the spectroscopic sample to test whether 
the [O~II] EW$_R$ is well correlated with the \ha \ EW$_R$.
We compare the \citet{postman98}  [O~II] EW$_R$ 
to our \ha \ EW$_R$
in  Figure \ref{sfrhao2} for the eleven
spectroscopically confirmed cluster members.  
The data are in good agreement
with the empirical relation for local galaxies \citep{kenn92a,kenn92b}.
The agreement between our \ha \ EW$_R$ and the 
spectroscopically determined [O~II] EW$_R$ provides further
confirmation of our imaging technique.

\subsection{Completeness of Spectroscopic Surveys}

The advantage of an imaging survey is that there is no analog to spectroscopic
target selection,
and therefore any object with significant star formation will be identified.
As a first indication that most of the star formation is 
missed in state-of-the-art
magnitude-limited spectroscopic surveys like that of \citet{postman98}, 
consider that we detect significant \ha\ emission from 35 galaxies, while 
\citet{postman98} detect [O II] emission from nine cluster galaxies
over the same area.
Some (1-5) of our detections may be non-cluster galaxies,
so we estimate that the spectroscopic
survey missed $\sim$70\% of the star-forming galaxies in this cluster
because of incomplete sampling.
In terms of total star-formation, we measure
an integrated SFR of 23.2$\pm$7.5~\smy \ for the \citeauthor{postman98} cluster galaxies,
and 67.8$\pm$20.5~\smy \  for our entire sample of 35 galaxies. 
Therefore, the \citet{postman98} survey misses $\sim$65\% of the cluster star 
formation in this survey area.

\subsection{Radial Distribution of SFRs}
\begin{figure}
\plotone{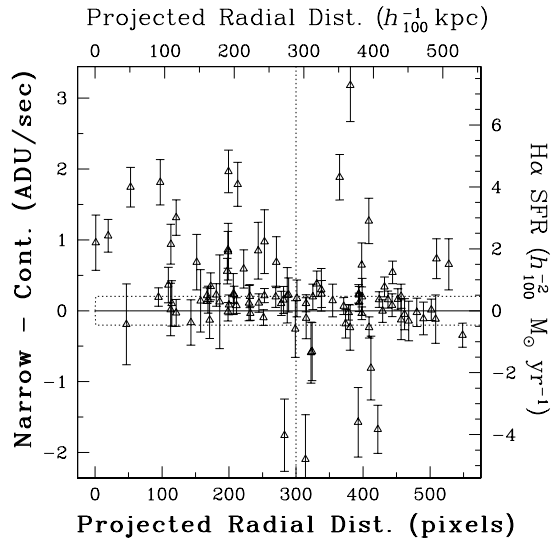}
\caption{Continuum-subtracted narrow-band flux versus projected
radial distance from cluster center for all galaxies in the \cj \ field.  
Horizontal dotted lines show 3$\sigma$ flux
limits.  Vertical dotted line marks radius beyond which areal
coverage is incomplete.}
\label{fd}
\end{figure}

In Figure \ref{fd} we plot the continuum-subtracted flux
versus projected radial distance from the cluster center
for all galaxies.
The cluster is not centrally concentrated, so its
center is not well-determined.  We define the center using 
the brightest cluster galaxy and find
no clear trend of SFR with radius.   Because
this cluster is not massive 
($\sigma = 415$~km~s$^{-1}$; \citeauthor{postman98} 
\citeyear{postman98}), 
it is likely to be quite irregular with a high fraction of emission line
galaxies even within the core.

To determine how well our data sample the radial extent
of the cluster and to define a radial benchmark for comparison
among clusters, we calculate $R_{200}$, which approximates
the virial radius.  $R_{200}$ is defined as the 
radius inside which the density is $200 \times$ the critical density:
\begin{equation}
200 \ \rho_c(z) = \frac{M_{cl}}{4/3 \pi R_{200}^3},
\end{equation}
Using the redshift dependence of the critical density and the
virial mass to relate the line-of-sight velocity dispersion, $\sigma_x$,
to the cluster mass, we express $R_{200}$ as
\begin{equation}
R_{200} = 1.73 \ \frac{\sigma_x}{1000~{\rm km/s}} \
[\Omega_\Lambda + \Omega_0 (1+z)^3]^{-\frac{1}{2}}\  h_{100}^{-1} \ {\rm Mpc}.
\end{equation}
Our areal coverage is complete
to 300~\h~kpc and stops at 550~\h~kpc,
yet $R_{200} \simeq 430$~\h~kpc so we are imaging
a significant fraction of this cluster.

\subsection{Morphological Dependence of SFRs}
\label{clj0023_morphs}

\begin{figure}
\plotone{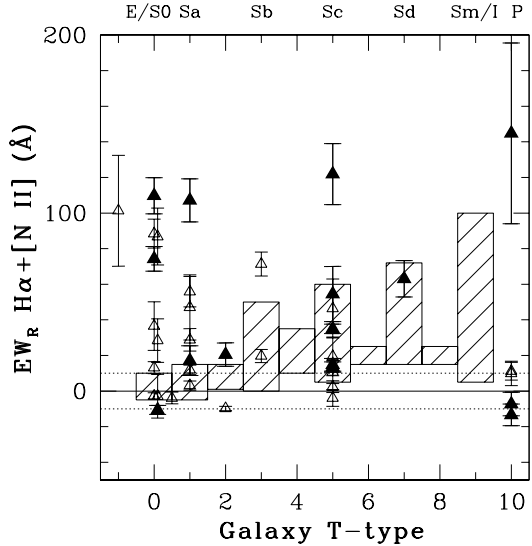}
\caption{EW$_R$ versus galaxy T-type for galaxies
with visually determined morphologies from \citet{lubin98a}.  We denote
an extremely compact source with a T-type of $-1$ and 
galaxies that \citeauthor{lubin98a} classify as peculiar with
a T-type of 10.  Galaxies with some sign of interaction are
shown with filled triangles.  Shaded areas show range
of EWs for sample of nearby galaxies \citep{kenn98}.}
\label{ewtype}
\end{figure}

The connection between morphology and SFR, which is
well established at low redshift 
\citep[e.g.][and references therein]{kenn98}, might be expected to
evolve at higher redshifts. Using visual morphological classifications
provided by 
\citet{lubin98a} from their WFPC2 imaging of 
the 200 brightest galaxies in the 
CL~J0023+0423B field, we begin to explore this relationship. 
In Figure \ref{ewtype}, we plot EW$_R$ versus galaxy
T-type for 39 galaxies that are either confirmed cluster members or 
galaxies with undetermined redshifts
(possible cluster members).  Our EW$_R$ detection
thresholds are delineated with horizontal dotted lines.  
The shaded boxes represent the range of observed equivalent widths for
nearby galaxies \citep{kenn98}.
At first glance, the striking result is that there are
several E/S0 galaxies with significant \ha\ EWs, 
which is not seen among the nearby galaxies \citep{kenn98}.
However, the morphological classifications were done visually, and we now
proceed to test this result by using an automated morphological
classification algorithm. 

\begin{figure}
\plotone{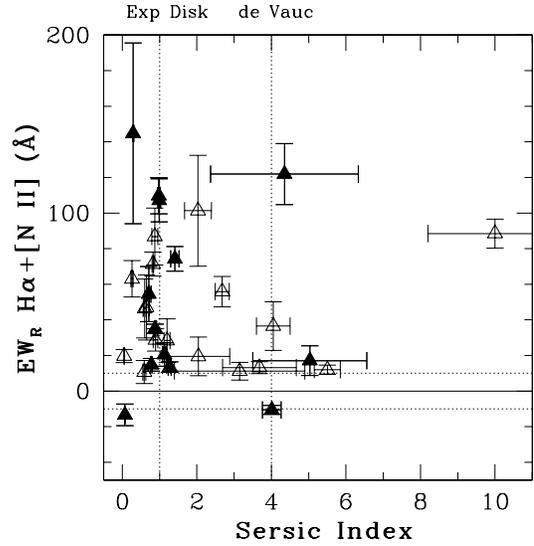}
\caption{EW$_R$ versus Sersic index calculated by
GALFIT \citep{peng02}.  
Galaxies with some sign of interaction are
shown with filled triangles.}
\label{ewsersic}
\end{figure}

\begin{figure}
\epsscale{.55}
\caption{(see 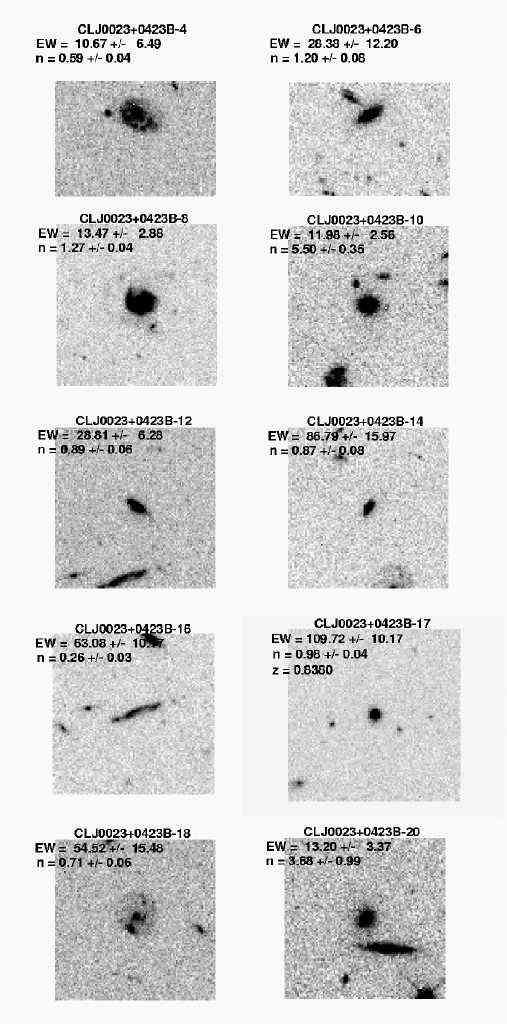) WFPC2 F702W $101\times 101$ pixel images of galaxies 
with EW$_R>10$~\AA.  The galaxy name,
narrow-band EW$_R$ (\AA), Sersic index, and redshift from \citet{postman98} 
if known 
are listed from top to bottom for each galaxy.
The two images taken with WFPC2 Planetary Camera are labeled 
PC in the top right corner.  All other images are taken with the wide-field
cameras.}
\label{mosaic}
\end{figure}

To obtain a more robust and uniform measure of the 
structural parameters of these galaxies, 
we use the GALFIT program \citep{peng02} to reanalyze the $HST$ WFPC2
F702W images. 
We fit Sersic profiles to the 30/39 galaxies in 
Figure \ref{ewtype} that have significant EWs and are therefore
likely cluster members.  The Sersic profile is given by 
\begin{equation}
\Sigma(r) = \Sigma_e e^{- \kappa[(r/r_e)^{\frac{1}{n}} -1]} ,
\end{equation}
where $r_e$ is the effective radius, $\Sigma_e$ is the surface brightness
at $r_e$, $n$ is the power-law index, and $\kappa$ is adjusted
so that half of the galaxy light lies within $r_e$ \citep{peng02}.
An exponential disk corresponds to $n=1$ while a 
de~Vaucouleur profile corresponds to $n=4$.  In Figure \ref{ewsersic}
we plot the EW$_R$ versus best-fit Sersic index.
In Figure \ref{mosaic} we present the 101$\times$101 
pixel WFPC2 F702W image of each galaxy that was supplied to GALFIT, the name 
(same as in Table 2), 
the EW$_R$, the
Sersic index, and the spectroscopic redshift if known.  The catalog name from 
\citet{lubin98a}, redshift \citep{postman98}, visual classification \citep{lubin98a}, 
and Sersic index 
are also listed in columns 13$-$16 of Table 2.

There are three interesting results from the GALFIT fitting.
First, 
four galaxies with significant emission that were visually classified 
as early-type (CLJ0023+0423B$-$6, 14, 17, and 76) are best fit by 
GALFIT with exponential disk profiles.
Therefore, their 
visual classification as early-type is suspect.
Second, several galaxies have best-fit Sersic indices greater than four.
In some cases the uncertainty of the index is quite high and reflects
GALFIT's difficulty in locating the galaxy center; visual inspection shows
these galaxies to be irregular.  We find one case where the 
high Sersic index is
relatively well constrained (CLJ0023+0423B$-$72), and this may indicate 
the presence of a central point source
or active galactic nucleus (AGN).  The lack of many such galaxies suggests 
that central point sources, dominant
in AGN, are rare. The issue of contamination of the 
line flux from AGN emission is
difficult to address without spectra, but even in those galaxies hosting AGN,
the contamination must be rather moderate given the moderate values of
$n$.  The third interesting feature of Figure \ref{ewsersic} is that
one galaxy with a Sersic index of four shows signs of ongoing
star-formation: CLJ0023+0423B$-$60 with an EW$_R$ of $36.5\pm 13.7$~\AA.
Inspection of the WFPC2 image shows a faint companion 
0.7\arcsec \ away from CLJ0023+0423B$-$60
that we do not resolve in our ground-based
data.  This companion could be contributing to the excess narrow-band flux.
This object warrants follow-up with near-infrared spectroscopy to determine
the redshift, confirm the \ha \ flux, and study the spectral
properties of the underlying stellar population.  
In conclusion, the results from the GALFIT analysis show that
EW$_R$ decreases with increasing steepness of the galaxy profile as expected
and that there is no significant population of early-type galaxies
with ongoing star-formation.

\begin{figure}
\epsscale{.55}
\begin{flushleft}
{Figure \ref{mosaic} --- Continued. (see 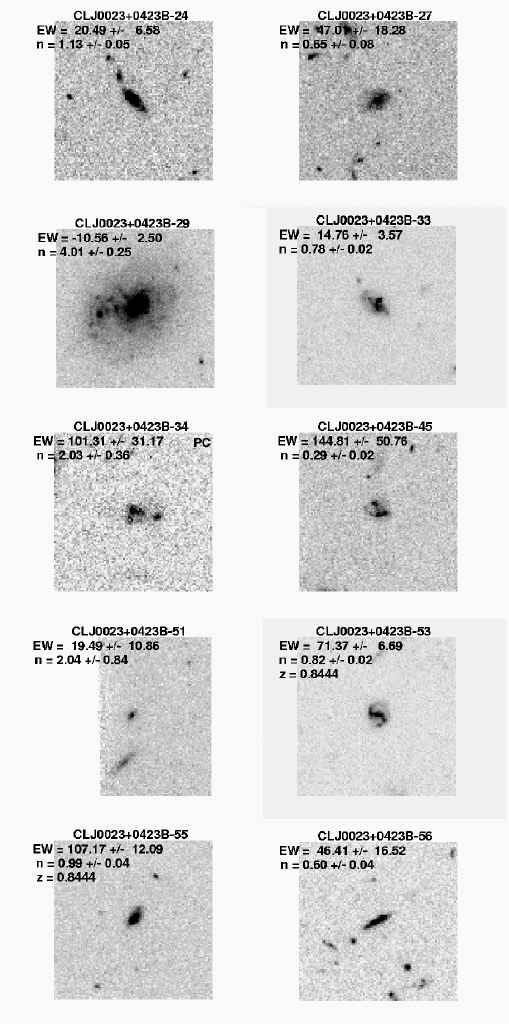)  
}
\end{flushleft}
\end{figure}

\begin{figure}
\epsscale{.55}
\begin{flushleft}
{Figure \ref{mosaic} --- Continued.  (see 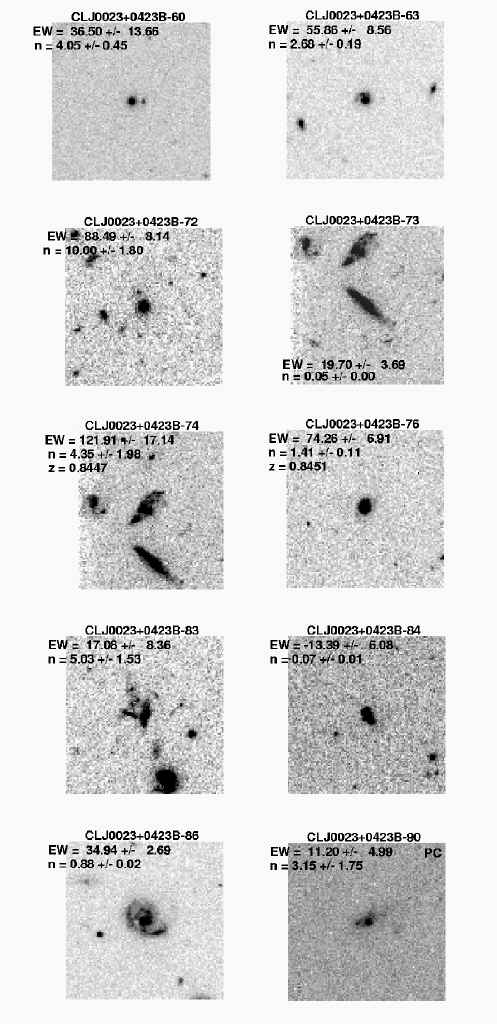)
}
\end{flushleft}
\end{figure}

\section{DISCUSSION}  
\label{clj0023_discussion}
In this section we compare the star-formation properties
of \cj \ with comparable observations for other clusters at lower redshifts.
Comparing spectroscopic and imaging surveys is difficult,
and comparing SFR measurements obtained with \ha \  vs.  
[O~II] is even more problematic.
Because of these difficulties, we limit our comparison 
to three $z \sim 0.2$ clusters
from the literature for which \ha \ imaging or spectroscopy is
available.  The three surveys are summarized in Table \ref{cljlitsfrs}. 

Analogous to the Butcher-Oemler blue fraction measurement
is the fraction of emission-line galaxies. 
To provide a fair comparison to the low-redshift \ha \ studies,
we have to apply four criteria to all clusters.  First,
we must sample the same radial fraction of each cluster relative
to $R_{200}$.  Then, we must apply the same EW, SFR, and 
absolute magnitude cuts to all data.  The low-redshift
spectroscopic samples are sensitive in terms of SFR and EW but have
limited radial coverage; the spectroscopic data for Abell~1689 extends to 
only $0.27 \times R_{200}$ \citep{balogh02}.  
The imaging survey of \citet{balogh00} has
good areal coverage relative to $R_{200}$ but is sensitive to 
only the most actively star-forming
galaxies with $EW > 50$~\AA.  
Thus, a fair comparison among the clusters in Table \ref{cljlitsfrs} requires
that we sample only within $0.27 \times R_{200}$ and include only
galaxies with $EW > 50$~\AA \ as emission line.
In addition, we must correct the non-emission line galaxy
counts because a large faction \citep[$\sim$50\%;][]{maihara01}
are not cluster members, and we do not have a large enough field area
to do statistical subtraction. 
The fraction of emission-line galaxies is therefore complicated to 
calculate and uses only a small fraction of the available data.

\begin{figure}
\epsscale{1}
\plotone{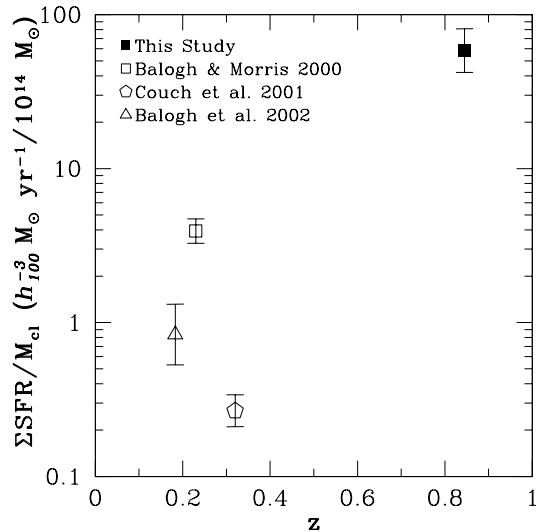}
\caption{Integrated SFR per cluster mass 
versus cluster redshift for \cj \ and 
three $z \sim 0.2$ clusters from the literature.  All
$>3\sigma$ sources within $0.5 \times R_{200}$ are 
included.
}
\label{clj0023_sfrz}
\end{figure}

\begin{figure}
\plotone{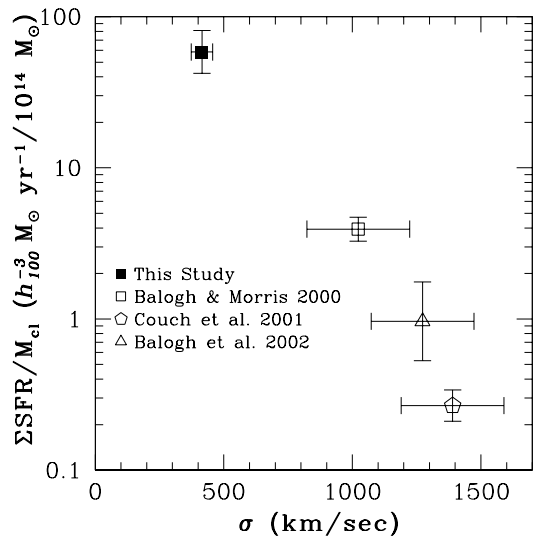}
\caption{Integrated cluster SFR per cluster mass versus cluster 
velocity dispersion for
CL~J0023+0423B and clusters from literature.  
All $>3\sigma$ sources within $0.5\times R_{200}$ are included.
}
\label{clj0023_sfrsigma}
\end{figure} 

Perhaps a simpler way to quantify the evolution of cluster SFRs is in terms
of the integrated SFR per cluster mass.
No correction for background/foreground galaxies is required, and 
this measure
is less sensitive to incompleteness because the integrated SFR is 
dominated by galaxies whose SFR is easiest to detect.  
To calculate, we apply the same radial cut to all four clusters, 
including only those galaxies that lie within a projected 
radial distance of $0.5 \times R_{200}$.  
We choose a maximum radial extent of $0.5\times R_{200}$
to approximate the radial survey size of AC~114 
\citep[see Table \ref{cljlitsfrs};][]{couch01}.
The radial coverage of Abell 1689 does not extend to 
$0.5 \times R_{200}$, so we multiply the integrated SFR 
by 1.35 to correct for incomplete sampling within $0.5 \times R_{200}$, 
where we assume that
the galaxy distribution follows the dark matter profile.
This correction will still underestimate the integrated SFR
if there is a strong increasing radial gradient in SFR.
The integrated SFR for Abell 1689 listed in Table \ref{cljlitsfrs} is corrected
for volume.
The integrated SFR per cluster mass might be expected to be a 
function of both redshift
and cluster mass (even beyond the normalization by cluster mass
one might expect that more massive clusters are more evolved), and in 
Figure \ref{clj0023_sfrz} 
we plot SFR versus redshift for the four clusters and versus cluster 
velocity dispersion in  Figure \ref{clj0023_sfrsigma}.
These plots exemplify a key difficulty in assessing the evolution of
cluster star formation, which is present in all studies of the 
Butcher-Oemler effect.
A larger sample of $z \sim 0.8$ clusters is needed to break this
degeneracy between mass and redshift.

\section{SUMMARY}
\label{clj0023_summary}
We present \ha-derived star-formation rates for the galaxy cluster 
CL~J0023+0423B ($z = 0.845$) and demonstrate 
that near-infrared narrow-band imaging
is an effective and efficient method to sample the star-forming 
galaxy population in distant clusters.
Comparison with spectroscopy shows that the number of false detections
is low ($9\pm 6$\%), and that our \ha \ equivalent widths 
are correlated with the
spectroscopically determined [O~II] equivalent widths.
We find that the magnitude-limited 
spectroscopic survey of this cluster by \citet{postman98}
misses $\sim$70\% of the star-forming galaxies and $\sim$65\% of the 
integrated star formation within our image area.
We fit Sersic profiles to $HST$ WFPC2 images of all galaxies
with significant EW$_R$ and find that EW$_R$ decreases as
the light profile steepens.  No significant population
of early-type galaxies with ongoing star formation is detected.
We compare \cj \ to three $z \sim 0.2$ clusters from the 
literature for which \ha \ data are available and find that
the integrated, mass-normalized SFR of \cj \ is a factor of ten
higher than that of the lower-redshift clusters.
Interpreting this difference is complicated by the strong correlation
between integrated SFR and cluster mass that we find.
Imaging for a larger sample of $z \approx 0.8$ clusters
drawn from the EDisCS  sample\footnote{http://www.mpa-garching.mpg.de/\~{}ediscs/project.html}, which
spans a range in cluster mass, is ongoing.

\acknowledgements
We thank Marc Postman for providing positional 
information of spectroscopic targets.  In addition, we thank 
John Moustakas, Rob Kennicutt, 
Stephane Charlot, and the referee James Rose for comments 
that improved the content and clarity of this paper. 
RAF acknowledges support from the NASA Graduate Student Researchers
Program through NASA Training Grant 
NGT5-50283
and from an NSF Astronomy and Astrophysics
Postdoctoral Fellowship under award AST-0301328.
DZ acknowledges support from the David and Lucile Packard Fellowship.
This research has made extensive use of the following: 
(1) NASA's Astrophysics Data System;  
(2) the 
NASA/IPAC Extragalactic
Database (NED) which is operated by the Jet Propulsion
Laboratory, California Institute of Technology, 
under contract with the National Aeronautics and Space Administration;
and (3) online data catalogs provided by the Centre Donn\'ees 
astronomiques de Strasbourg.


\clearpage
\begin{deluxetable}{cccccc} 
\tablecaption{Summary of Photometric Calibrations\label{phot}}
\tablehead{
&   \multicolumn{3}{c}{$J$-band Solution} & \multicolumn{2}{c}{Narrow-band ZP} \\
Date & ZP & Airmass Coeff & R.M.S & Flux\tablenotemark{a} &  SFR\tablenotemark{b}}
\startdata  

{2002 Dec 19$^c$} 	 &24.94$\pm$0.02 	& 0.06$\pm$0.01 	& 0.0400 & 9.21$\pm$0.06 & 2.29$\pm$0.02 \\
\enddata
\tablenotetext{a}{Units of $\rm 10^{-17}~ergs/ s/cm^{2}$.}
\tablenotetext{b}{SFR in units of \smy \ corresponding to a $z = 0.845$ source.}
\tablenotetext{c}{Solution from standard stars P525-E, P533-D, and S840-F \citep{persson98}. }
\end{deluxetable}

\begin{turnpage}
\begin{deluxetable}{rrrrrrrrrrrrrrrr} 
\tabletypesize{\tiny}
\tablecaption{\ha \ Data for CL J0023+0423B Galaxies \label{clj0023sfrtable}} 
\tablehead{\colhead{Name} & \colhead{$\delta$RA} & \colhead{$\delta$Dec} & \colhead{x} & \colhead{y} & \colhead{Flux$_n$} & \colhead{Flux$_J$} & \colhead{Cont. Sub} & \colhead{EW$_R$(\ha)} & \colhead{Flux(\ha)} & \colhead{L(\ha)} & \colhead{SFR} & \colhead{MDS} & \colhead{z} & \colhead{T} & \colhead{n}\\ \colhead{(1)} &  \colhead{(2)} &  \colhead{(3)} &  \colhead{(4)} &  \colhead{(5)} &  \colhead{(6)} &  \colhead{(7)} & \colhead{(8)} &  \colhead{(9)} &  \colhead{(10)} &  \colhead{(11)} &  \colhead{(12)} & \colhead{(13)} & \colhead{(14)} & \colhead{(15)}& \colhead{(16)}}
\startdata 
1 &    -71.2 &    -41.1 &    326.5 &     21.3 &      1.5 $\pm$      0.1 &     33.9 $\pm$      1.1  &     -0.1 $\pm$      0.1  &  -5.9 $\pm$   5.5 & -1.13 $\pm$  1.07 &  -0.4 $\pm$   0.3 &  -0.3 $\pm$   0.3 & \ldots & \ldots &  \ldots & \ldots\\ 
2 &    -72.1 &    -24.1 &    421.1 &     16.6 &      3.3 $\pm$      0.1 &    100.2 $\pm$      1.4  &     -1.6 $\pm$      0.2  & -26.1 $\pm$   2.2 & -14.82 $\pm$  1.30 &  -4.9 $\pm$   0.4 &  -3.8 $\pm$   1.2 & \ldots & \ldots &  \ldots & \ldots\\ 
3 &    -62.1 &     33.7 &    742.2 &     71.8 &     11.1 $\pm$      0.2 &    258.6 $\pm$      1.9  &     -1.5 $\pm$      0.2  &  -9.5 $\pm$   1.3 & -13.93 $\pm$  1.91 &  -4.6 $\pm$   0.6 &  -3.6 $\pm$   1.2 & \ldots & \ldots &  \ldots & \ldots\\ 
4 &    -64.3 &     31.7 &    731.2 &     60.0 &      3.3 $\pm$      0.1 &     55.5 $\pm$      1.2  &      0.6 $\pm$      0.1  &  18.2 $\pm$   4.0 &  5.71 $\pm$  1.21 &   1.9 $\pm$   0.4 &   1.5 $\pm$   0.5 &  26  & \ldots   &  5        &    0.59 $\pm$   0.04 \\ 
5 &    -55.1 &     19.1 &    661.1 &    111.0 &      5.8 $\pm$      0.2 &    129.9 $\pm$      1.6  &     -0.6 $\pm$      0.2  &  -6.9 $\pm$   2.1 & -5.10 $\pm$  1.58 &  -1.7 $\pm$   0.5 &  -1.3 $\pm$   0.6 & \ldots & \ldots &  \ldots & \ldots\\ 
6 &    -59.6 &    -12.4 &    486.0 &     86.0 &      1.0 $\pm$      0.1 &     16.1 $\pm$      0.8  &      0.2 $\pm$      0.1  &  23.0 $\pm$   8.8 &  2.09 $\pm$  0.74 &   0.7 $\pm$   0.2 &   0.5 $\pm$   0.3 &  32  & \ldots    &  0.1      &    1.20 $\pm$   0.08 \\ 
7 &    -57.0 &    -43.6 &    312.7 &    100.4 &      1.6 $\pm$      0.1 &     28.7 $\pm$      1.1  &      0.2 $\pm$      0.1  &  10.7 $\pm$   6.5 &  1.73 $\pm$  1.02 &   0.6 $\pm$   0.3 &   0.4 $\pm$   0.3 & \ldots & \ldots &  \ldots & \ldots\\ 
8 &    -57.6 &    -15.2 &    470.7 &     96.8 &      0.9 $\pm$      0.1 &     10.1 $\pm$      0.7  &      0.4 $\pm$      0.1  &  58.9 $\pm$  14.6 &  3.35 $\pm$  0.69 &   1.1 $\pm$   0.2 &   0.9 $\pm$   0.3 &  15  & \ldots   &  5        &    1.27 $\pm$   0.04 \\ 
9 &    -53.6 &    -23.6 &    424.1 &    119.5 &      0.7 $\pm$      0.1 &     10.8 $\pm$      0.7  &      0.2 $\pm$      0.1  &  28.4 $\pm$  12.2 &  1.73 $\pm$  0.68 &   0.6 $\pm$   0.2 &   0.4 $\pm$   0.2 & \ldots & \ldots &  \ldots & \ldots\\ 
10 &    -51.9 &    -69.1 &    170.9 &    128.5 &      0.8 $\pm$      0.1 &     17.1 $\pm$      0.8  &     -0.0 $\pm$      0.1  &  -2.2 $\pm$   7.8 & -0.21 $\pm$  0.76 &  -0.1 $\pm$   0.2 &  -0.1 $\pm$   0.2 &  24  & \ldots	 &  1        &    5.50 $\pm$   0.35 \\ 
11 &    -42.5 &    -10.5 &    496.4 &    180.7 &      5.6 $\pm$      0.1 &     98.1 $\pm$      1.6  &      0.8 $\pm$      0.2  &  13.5 $\pm$   2.9 &  7.49 $\pm$  1.54 &   2.5 $\pm$   0.5 &   1.9 $\pm$   0.7 & \ldots & \ldots &  \ldots & \ldots\\ 
12 &    -39.3 &     13.0 &    627.5 &    198.9 &      1.5 $\pm$      0.1 &     27.8 $\pm$      1.0  &      0.1 $\pm$      0.1  &   6.2 $\pm$   6.2 &  0.97 $\pm$  0.96 &   0.3 $\pm$   0.3 &   0.3 $\pm$   0.3 &  47  & \ldots	 &  1        &    0.89 $\pm$   0.06 \\ 
13 &    -35.7 &      1.7 &    564.5 &    218.5 &      6.1 $\pm$      0.1 &    109.0 $\pm$      1.5  &      0.8 $\pm$      0.2  &  12.0 $\pm$   2.6 &  7.40 $\pm$  1.53 &   2.4 $\pm$   0.5 &   1.9 $\pm$   0.7 & \ldots & \ldots &  \ldots & \ldots\\ 
14 &    -38.0 &    -81.9 &     99.8 &    205.8 &      0.5 $\pm$      0.1 &     10.5 $\pm$      0.6  &      0.0 $\pm$      0.1  &   2.1 $\pm$   9.7 &  0.13 $\pm$  0.57 &   0.0 $\pm$   0.2 &   0.0 $\pm$   0.1 &  77  & \ldots	 &  0.1      &    0.87 $\pm$   0.08 \\ 
15 &    -36.5 &    -16.2 &    465.1 &    214.0 &      2.1 $\pm$      0.1 &     32.1 $\pm$      1.1  &      0.6 $\pm$      0.1  &  28.8 $\pm$   6.3 &  5.23 $\pm$  1.05 &   1.7 $\pm$   0.3 &   1.4 $\pm$   0.5 & \ldots & \ldots &  \ldots & \ldots\\ 
16 &    -31.8 &    -39.7 &    334.2 &    240.4 &     12.1 $\pm$      0.2 &    281.6 $\pm$      1.9  &     -1.7 $\pm$      0.2  &  -9.7 $\pm$   1.2 & -15.55 $\pm$  1.98 &  -5.1 $\pm$   0.6 &  -4.0 $\pm$   1.3 &  58  & \ldots	 &  7        &    0.26 $\pm$   0.03 \\ 
17 &    -35.1 &     -5.2 &    526.3 &    221.8 &      1.0 $\pm$      0.1 &     10.0 $\pm$      0.7  &      0.5 $\pm$      0.1  &  86.8 $\pm$  16.0 &  4.94 $\pm$  0.70 &   1.6 $\pm$   0.2 &   1.3 $\pm$   0.4 &  57  & 0.8380	 &  0        &    0.98 $\pm$   0.04 \\ 
18 &    -32.4 &    -60.4 &    219.6 &    236.9 &      3.3 $\pm$      0.1 &     72.6 $\pm$      1.3  &     -0.2 $\pm$      0.1  &  -5.1 $\pm$   3.0 & -2.11 $\pm$  1.24 &  -0.7 $\pm$   0.4 &  -0.5 $\pm$   0.4 &  48  & \ldots	 &  5        &    0.71 $\pm$   0.06 \\ 
19 &    -32.0 &    -15.9 &    466.5 &    239.5 &      1.9 $\pm$      0.1 &     21.3 $\pm$      1.0  &      0.8 $\pm$      0.1  &  63.1 $\pm$  10.2 &  7.62 $\pm$  1.01 &   2.5 $\pm$   0.3 &   2.0 $\pm$   0.6 & \ldots & \ldots &  \ldots & \ldots\\ 
20 &    -31.6 &     16.9 &    648.7 &    241.3 &      3.3 $\pm$      0.1 &     28.0 $\pm$      1.2  &      1.9 $\pm$      0.2  & 109.7 $\pm$  10.2 & 17.41 $\pm$  1.16 &   5.7 $\pm$   0.4 &   4.5 $\pm$   1.4 &  40  & \ldots	 &  0        &    3.68 $\pm$   0.99 \\ 
21 &    -30.2 &     -7.3 &    514.2 &    249.3 &      0.8 $\pm$      0.1 &      9.9 $\pm$      0.8  &      0.3 $\pm$      0.1  &  54.5 $\pm$  15.5 &  3.05 $\pm$  0.73 &   1.0 $\pm$   0.2 &   0.8 $\pm$   0.3 & \ldots & \ldots &  \ldots & \ldots\\ 
22 &    -30.7 &      2.8 &    570.7 &    246.6 &      1.4 $\pm$      0.1 &     30.8 $\pm$      1.1  &     -0.1 $\pm$      0.1  &  -6.7 $\pm$   5.7 & -1.17 $\pm$  1.01 &  -0.4 $\pm$   0.3 &  -0.3 $\pm$   0.3 & \ldots & \ldots &  \ldots & \ldots\\ 
23 &    -20.3 &     18.3 &    656.6 &    304.3 &      4.6 $\pm$      0.1 &     81.5 $\pm$      1.5  &      0.7 $\pm$      0.2  &  13.2 $\pm$   3.4 &  6.09 $\pm$  1.50 &   2.0 $\pm$   0.5 &   1.6 $\pm$   0.6 & \ldots & \ldots &  \ldots & \ldots\\ 
24 &    -27.6 &    -75.0 &    138.2 &    263.9 &      0.7 $\pm$      0.1 &      4.2 $\pm$      0.6  &      0.5 $\pm$      0.1  & 201.1 $\pm$  46.9 &  4.78 $\pm$  0.61 &   1.6 $\pm$   0.2 &   1.2 $\pm$   0.4 &  62  & \ldots	 &  2        &    1.13 $\pm$   0.05 \\ 
25 &    -17.7 &    -10.0 &    499.7 &    318.7 &      3.6 $\pm$      0.1 &     72.8 $\pm$      1.5  &      0.0 $\pm$      0.2  &   0.4 $\pm$   3.4 &  0.15 $\pm$  1.41 &   0.1 $\pm$   0.5 &   0.0 $\pm$   0.4 & \ldots & \ldots &  \ldots & \ldots\\ 
26 &    -22.6 &    -70.1 &    165.7 &    291.2 &      0.4 $\pm$      0.1 &     12.4 $\pm$      0.6  &     -0.2 $\pm$      0.1  & -29.6 $\pm$   7.8 & -2.08 $\pm$  0.59 &  -0.7 $\pm$   0.2 &  -0.5 $\pm$   0.2 & \ldots & \ldots &  \ldots & \ldots\\ 
27 &    -19.7 &     -0.6 &    551.8 &    307.6 &      1.7 $\pm$      0.1 &     27.6 $\pm$      1.0  &      0.3 $\pm$      0.1  &  20.5 $\pm$   6.6 &  3.20 $\pm$  0.97 &   1.0 $\pm$   0.3 &   0.8 $\pm$   0.4 &  71  & \ldots	 &  1        &    0.65 $\pm$   0.08 \\ 
28 &    -19.9 &     40.6 &    780.8 &    306.2 &      0.2 $\pm$      0.0 &      5.5 $\pm$      0.4  &     -0.1 $\pm$      0.0  & -27.8 $\pm$  12.2 & -0.86 $\pm$  0.41 &  -0.3 $\pm$   0.1 &  -0.2 $\pm$   0.1 & \ldots & \ldots &  \ldots & \ldots\\ 
29 &    -17.9 &    -30.9 &    383.4 &    317.5 &      0.3 $\pm$      0.0 &      7.2 $\pm$      0.5  &     -0.0 $\pm$      0.1  &  -2.9 $\pm$  12.2 & -0.12 $\pm$  0.50 &  -0.0 $\pm$   0.2 &  -0.0 $\pm$   0.1 &  7  & \ldots	 &  0.1      &    4.01 $\pm$   0.25 \\ 
30 &    -16.1 &    -38.4 &    341.9 &    327.6 &      0.5 $\pm$      0.1 &      6.8 $\pm$      0.6  &      0.2 $\pm$      0.1  &  47.0 $\pm$  18.3 &  1.81 $\pm$  0.61 &   0.6 $\pm$   0.2 &   0.5 $\pm$   0.2 & \ldots & \ldots &  \ldots & \ldots\\ 
31 &     44.5 &    -26.2 &    409.4 &    664.2 &      5.5 $\pm$      0.1 &    108.1 $\pm$      1.5  &      0.2 $\pm$      0.2  &   3.2 $\pm$   2.5 &  1.93 $\pm$  1.52 &   0.6 $\pm$   0.5 &   0.5 $\pm$   0.4 & \ldots & \ldots &  \ldots & \ldots\\ 
32 &     70.6 &    -22.6 &    429.5 &    809.3 &      5.1 $\pm$      0.2 &    119.5 $\pm$      1.8  &     -0.8 $\pm$      0.2  & -10.6 $\pm$   2.5 & -7.15 $\pm$  1.74 &  -2.3 $\pm$   0.6 &  -1.8 $\pm$   0.7 & \ldots & \ldots &  \ldots & \ldots\\ 
33 &     69.7 &     13.4 &    629.2 &    804.0 &      0.4 $\pm$      0.0 &      3.0 $\pm$      0.5  &      0.2 $\pm$      0.1  & 115.7 $\pm$  38.1 &  1.97 $\pm$  0.45 &   0.6 $\pm$   0.1 &   0.5 $\pm$   0.2 &  27  & \ldots	 &  5        &    0.78 $\pm$   0.02 \\ 
34 &     65.8 &    -43.2 &    315.3 &    782.5 &      0.4 $\pm$      0.0 &      4.2 $\pm$      0.5  &      0.1 $\pm$      0.1  &  57.4 $\pm$  23.0 &  1.38 $\pm$  0.46 &   0.5 $\pm$   0.2 &   0.4 $\pm$   0.2 & 120  & \ldots	 &  -1       &    2.03 $\pm$   0.36 \\ 
35 &     39.2 &    -14.4 &    474.9 &    634.7 &      0.2 $\pm$      0.0 &      4.7 $\pm$      0.4  &     -0.0 $\pm$      0.0  & -12.3 $\pm$  14.5 & -0.33 $\pm$  0.40 &  -0.1 $\pm$   0.1 &  -0.1 $\pm$   0.1 & \ldots & \ldots &  \ldots & \ldots\\ 
36 &     35.7 &    -33.2 &    370.7 &    615.1 &      4.2 $\pm$      0.1 &     72.6 $\pm$      1.4  &      0.7 $\pm$      0.2  &  14.8 $\pm$   3.6 &  6.07 $\pm$  1.41 &   2.0 $\pm$   0.5 &   1.6 $\pm$   0.6 & \ldots & \ldots &  \ldots & \ldots\\ 
37 &     36.4 &      8.6 &    602.6 &    619.2 &      0.4 $\pm$      0.0 &      3.5 $\pm$      0.5  &      0.2 $\pm$      0.1  & 101.3 $\pm$  31.2 &  2.03 $\pm$  0.45 &   0.7 $\pm$   0.1 &   0.5 $\pm$   0.2 & \ldots & \ldots &  \ldots & \ldots\\ 
38 &     35.5 &    -84.6 &     85.1 &    614.4 &      2.5 $\pm$      0.1 &     37.5 $\pm$      1.1  &      0.7 $\pm$      0.1  &  30.6 $\pm$   5.7 &  6.51 $\pm$  1.10 &   2.1 $\pm$   0.4 &   1.7 $\pm$   0.6 & \ldots & \ldots &  \ldots & \ldots\\ 
39 &     35.5 &    -74.2 &    143.0 &    614.3 &      0.8 $\pm$      0.1 &     11.3 $\pm$      0.7  &      0.2 $\pm$      0.1  &  28.4 $\pm$  11.1 &  1.82 $\pm$  0.66 &   0.6 $\pm$   0.2 &   0.5 $\pm$   0.2 & \ldots & \ldots &  \ldots & \ldots\\ 
40 &     35.3 &    -75.3 &    136.7 &    613.3 &      1.1 $\pm$      0.1 &     23.0 $\pm$      0.9  &     -0.0 $\pm$      0.1  &  -3.1 $\pm$   6.5 & -0.41 $\pm$  0.86 &  -0.1 $\pm$   0.3 &  -0.1 $\pm$   0.2 & \ldots & \ldots &  \ldots & \ldots\\ 
41 &     33.5 &     -0.1 &    554.3 &    602.9 &     18.7 $\pm$      0.2 &    380.0 $\pm$      2.5  &      0.1 $\pm$      0.3  &   0.5 $\pm$   1.2 &  1.14 $\pm$  2.53 &   0.4 $\pm$   0.8 &   0.3 $\pm$   0.7 & \ldots & \ldots &  \ldots & \ldots\\ 
42 &     34.0 &    -41.7 &    323.4 &    606.0 &      6.9 $\pm$      0.2 &    146.5 $\pm$      1.5  &     -0.2 $\pm$      0.2  &  -2.7 $\pm$   1.9 & -2.28 $\pm$  1.56 &  -0.7 $\pm$   0.5 &  -0.6 $\pm$   0.4 & \ldots & \ldots &  \ldots & \ldots\\ 
43 &     30.9 &    -73.6 &    146.2 &    588.9 &      0.6 $\pm$      0.1 &      9.7 $\pm$      0.6  &      0.1 $\pm$      0.1  &  12.5 $\pm$  11.3 &  0.69 $\pm$  0.60 &   0.2 $\pm$   0.2 &   0.2 $\pm$   0.2 & \ldots & \ldots &  \ldots & \ldots\\ 
44 &     30.3 &    -71.6 &    157.0 &    585.2 &      0.5 $\pm$      0.1 &      3.4 $\pm$      0.5  &      0.3 $\pm$      0.1  & 156.3 $\pm$  44.7 &  2.99 $\pm$  0.52 &   1.0 $\pm$   0.2 &   0.8 $\pm$   0.3 & \ldots & \ldots &  \ldots & \ldots\\ 
45 &     29.6 &    -78.9 &    116.6 &    581.4 &      2.6 $\pm$      0.1 &     56.0 $\pm$      1.1  &     -0.1 $\pm$      0.1  &  -4.0 $\pm$   3.4 & -1.27 $\pm$  1.10 &  -0.4 $\pm$   0.4 &  -0.3 $\pm$   0.3 &  69  & \ldots	 &  10       &    0.29 $\pm$   0.02 \\ 
46 &     31.2 &    -93.7 &     34.5 &    590.2 &      0.4 $\pm$      0.1 &     14.9 $\pm$      0.7  &     -0.3 $\pm$      0.1  & -36.3 $\pm$   7.0 & -3.06 $\pm$  0.64 &  -1.0 $\pm$   0.2 &  -0.8 $\pm$   0.3 & \ldots & \ldots &  \ldots & \ldots\\ 
47 &     29.0 &    -86.8 &     73.0 &    578.3 &      4.0 $\pm$      0.1 &     83.4 $\pm$      1.3  &     -0.1 $\pm$      0.1  &  -2.2 $\pm$   2.8 & -1.05 $\pm$  1.33 &  -0.3 $\pm$   0.4 &  -0.3 $\pm$   0.4 & \ldots & \ldots &  \ldots & \ldots\\ 
48 &     27.5 &    -12.3 &    486.8 &    569.6 &      0.3 $\pm$      0.0 &      2.3 $\pm$      0.4  &      0.2 $\pm$      0.0  & 144.8 $\pm$  50.8 &  1.91 $\pm$  0.42 &   0.6 $\pm$   0.1 &   0.5 $\pm$   0.2 & \ldots & \ldots &  \ldots & \ldots\\ 
49 &     19.3 &    -69.0 &    171.6 &    524.5 &      0.2 $\pm$      0.0 &      4.0 $\pm$      0.4  &     -0.0 $\pm$      0.0  & -14.3 $\pm$  15.4 & -0.33 $\pm$  0.37 &  -0.1 $\pm$   0.1 &  -0.1 $\pm$   0.1 & \ldots & \ldots &  \ldots & \ldots\\ 
50 &     21.1 &     -5.4 &    525.0 &    534.3 &      0.7 $\pm$      0.1 &     15.9 $\pm$      0.8  &     -0.0 $\pm$      0.1  &  -3.0 $\pm$   8.2 & -0.27 $\pm$  0.75 &  -0.1 $\pm$   0.2 &  -0.1 $\pm$   0.2 & \ldots & \ldots &  \ldots & \ldots\\ 
\enddata
\end{deluxetable}
\setcounter{table}{1}
\begin{deluxetable}{rrrrrrrrrrrrrrrr} 
\tabletypesize{\tiny}
\tablecaption{---  Cont'd} 
\tablehead{\colhead{Name} & \colhead{$\delta$RA} & \colhead{$\delta$Dec} & \colhead{x} & \colhead{y} & \colhead{Flux$_n$} & \colhead{Flux$_J$} & \colhead{Cont. Sub} & \colhead{EW$_R$(\ha)} & \colhead{Flux(\ha)} & \colhead{L(\ha)} & \colhead{SFR} & \colhead{MDS} & \colhead{z} & \colhead{T} & \colhead{n}\\ \colhead{(1)} &  \colhead{(2)} &  \colhead{(3)} &  \colhead{(4)} &  \colhead{(5)} &  \colhead{(6)} &  \colhead{(7)} & \colhead{(8)} &  \colhead{(9)} &  \colhead{(10)} &  \colhead{(11)} &  \colhead{(12)} & \colhead{(13)} & \colhead{(14)} & \colhead{(15)}& \colhead{(16)}}
\startdata

51 &     25.0 &     -6.2 &    520.5 &    555.9 &      3.1 $\pm$      0.1 &     66.9 $\pm$      1.3  &     -0.2 $\pm$      0.1  &  -3.8 $\pm$   3.2 & -1.46 $\pm$  1.24 &  -0.5 $\pm$   0.4 &  -0.4 $\pm$   0.3 & 148  & \ldots	 &  5        &    2.04 $\pm$   0.84 \\ 
52 &     25.6 &     12.3 &    623.1 &    559.1 &      7.4 $\pm$      0.2 &    149.9 $\pm$      1.7  &      0.1 $\pm$      0.2  &   1.4 $\pm$   2.0 &  1.23 $\pm$  1.72 &   0.4 $\pm$   0.6 &   0.3 $\pm$   0.5 & \ldots & \ldots &  \ldots & \ldots\\ 
53 &     48.8 &    -58.7 &    228.7 &    687.9 &      0.3 $\pm$      0.0 &      2.6 $\pm$      0.4  &      0.2 $\pm$      0.0  &  93.5 $\pm$  36.5 &  1.40 $\pm$  0.41 &   0.5 $\pm$   0.1 &   0.4 $\pm$   0.2 &  30  & 0.8444	 &  3        &    0.82 $\pm$   0.02 \\ 
54 &     21.1 &    -21.8 &    433.8 &    534.4 &      0.8 $\pm$      0.1 &     12.5 $\pm$      0.8  &      0.2 $\pm$      0.1  &  19.5 $\pm$  10.9 &  1.38 $\pm$  0.73 &   0.5 $\pm$   0.2 &   0.4 $\pm$   0.2 & \ldots & \ldots &  \ldots & \ldots\\ 
55 &     45.7 &     33.5 &    741.3 &    670.9 &      0.3 $\pm$      0.0 &      4.5 $\pm$      0.5  &      0.1 $\pm$      0.0  &  35.6 $\pm$  19.4 &  0.91 $\pm$  0.45 &   0.3 $\pm$   0.1 &   0.2 $\pm$   0.1 &  50  & 0.8444	 &  1        &    0.99 $\pm$   0.04 \\ 
56 &     14.6 &      9.6 &    608.6 &    498.2 &      3.7 $\pm$      0.1 &     39.7 $\pm$      1.2  &      1.7 $\pm$      0.2  &  71.4 $\pm$   6.7 & 16.06 $\pm$  1.22 &   5.3 $\pm$   0.4 &   4.2 $\pm$   1.3 &  60  & \ldots	 &  5        &    0.60 $\pm$   0.04 \\ 
57 &     49.6 &     40.2 &    778.3 &    692.6 &      1.3 $\pm$      0.1 &     23.0 $\pm$      0.9  &      0.1 $\pm$      0.1  &   9.9 $\pm$   6.9 &  1.29 $\pm$  0.87 &   0.4 $\pm$   0.3 &   0.3 $\pm$   0.2 & \ldots & \ldots &  \ldots & \ldots\\ 
58 &     15.0 &    -15.8 &    467.2 &    500.6 &      2.2 $\pm$      0.1 &     19.2 $\pm$      0.9  &      1.3 $\pm$      0.1  & 107.2 $\pm$  12.1 & 11.64 $\pm$  0.95 &   3.8 $\pm$   0.3 &   3.0 $\pm$   0.9 & \ldots & \ldots &  \ldots & \ldots\\ 
59 &     14.5 &     29.2 &    717.2 &    497.5 &      0.6 $\pm$      0.1 &      7.5 $\pm$      0.6  &      0.2 $\pm$      0.1  &  46.4 $\pm$  16.5 &  1.98 $\pm$  0.61 &   0.6 $\pm$   0.2 &   0.5 $\pm$   0.2 & \ldots & \ldots &  \ldots & \ldots\\ 
60 &     14.0 &     15.1 &    638.9 &    494.8 &      3.4 $\pm$      0.1 &     66.6 $\pm$      1.2  &      0.1 $\pm$      0.1  &   2.5 $\pm$   3.3 &  0.94 $\pm$  1.22 &   0.3 $\pm$   0.4 &   0.2 $\pm$   0.3 &  85  & \ldots	 &  0        &    4.05 $\pm$   0.45 \\ 
61 &     11.5 &    -55.5 &    246.5 &    481.1 &      1.9 $\pm$      0.1 &     42.0 $\pm$      1.2  &     -0.1 $\pm$      0.1  &  -3.9 $\pm$   4.6 & -0.93 $\pm$  1.11 &  -0.3 $\pm$   0.4 &  -0.2 $\pm$   0.3 & \ldots & \ldots &  \ldots & \ldots\\ 
62 &     16.1 &     -5.7 &    523.6 &    506.4 &      0.4 $\pm$      0.0 &      3.9 $\pm$      0.5  &      0.2 $\pm$      0.1  &  77.2 $\pm$  28.5 &  1.71 $\pm$  0.49 &   0.6 $\pm$   0.2 &   0.4 $\pm$   0.2 & \ldots & \ldots &  \ldots & \ldots\\ 
63 &     23.2 &    -42.7 &    317.9 &    545.7 &      0.6 $\pm$      0.1 &      8.5 $\pm$      0.6  &      0.2 $\pm$      0.1  &  36.5 $\pm$  13.7 &  1.75 $\pm$  0.59 &   0.6 $\pm$   0.2 &   0.5 $\pm$   0.2 &  56  & \ldots	 &  1        &    2.68 $\pm$   0.19 \\ 
64 &     36.8 &      5.2 &    584.0 &    621.5 &      1.9 $\pm$      0.1 &     34.6 $\pm$      1.2  &      0.2 $\pm$      0.1  &  10.7 $\pm$   5.9 &  2.10 $\pm$  1.12 &   0.7 $\pm$   0.4 &   0.5 $\pm$   0.3 & \ldots & \ldots &  \ldots & \ldots\\ 
65 &      8.2 &    -56.0 &    244.2 &    462.6 &     18.8 $\pm$      0.2 &    427.2 $\pm$      2.3  &     -2.0 $\pm$      0.3  &  -7.7 $\pm$   1.0 & -18.58 $\pm$  2.43 &  -6.1 $\pm$   0.8 &  -4.8 $\pm$   1.6 & \ldots & \ldots &  \ldots & \ldots\\ 
66 &     10.5 &     17.5 &    652.0 &    475.2 &      2.2 $\pm$      0.1 &     26.3 $\pm$      1.1  &      0.9 $\pm$      0.1  &  55.9 $\pm$   8.6 &  8.31 $\pm$  1.07 &   2.7 $\pm$   0.4 &   2.1 $\pm$   0.7 & \ldots & \ldots &  \ldots & \ldots\\ 
67 &     45.6 &    -29.3 &    392.4 &    670.5 &      2.1 $\pm$      0.1 &     38.9 $\pm$      1.0  &      0.2 $\pm$      0.1  &   7.1 $\pm$   4.7 &  1.56 $\pm$  1.02 &   0.5 $\pm$   0.3 &   0.4 $\pm$   0.3 & \ldots & \ldots &  \ldots & \ldots\\ 
68 &     45.0 &      6.9 &    593.2 &    667.1 &      8.0 $\pm$      0.2 &    145.0 $\pm$      1.7  &      0.9 $\pm$      0.2  &  10.5 $\pm$   2.2 &  8.61 $\pm$  1.76 &   2.8 $\pm$   0.6 &   2.2 $\pm$   0.8 & \ldots & \ldots &  \ldots & \ldots\\ 
69 &     10.7 &    -76.5 &    130.0 &    476.4 &      0.5 $\pm$      0.1 &     10.4 $\pm$      0.7  &     -0.0 $\pm$      0.1  &  -0.8 $\pm$  10.5 & -0.05 $\pm$  0.62 &  -0.0 $\pm$   0.2 &  -0.0 $\pm$   0.2 & \ldots & \ldots &  \ldots & \ldots\\ 
70 &      9.6 &    -67.5 &    180.2 &    470.4 &      0.2 $\pm$      0.0 &      5.5 $\pm$      0.5  &     -0.0 $\pm$      0.0  &  -5.9 $\pm$  13.8 & -0.18 $\pm$  0.44 &  -0.1 $\pm$   0.1 &  -0.0 $\pm$   0.1 & \ldots & \ldots &  \ldots & \ldots\\ 
71 &      8.0 &     -2.6 &    540.3 &    461.6 &     12.6 $\pm$      0.2 &    262.1 $\pm$      2.2  &     -0.2 $\pm$      0.2  &  -1.2 $\pm$   1.5 & -1.72 $\pm$  2.22 &  -0.6 $\pm$   0.7 &  -0.4 $\pm$   0.6 & \ldots & \ldots &  \ldots & \ldots\\ 
72 &      8.9 &     36.9 &    760.2 &    466.6 &      0.3 $\pm$      0.0 &      4.2 $\pm$      0.5  &      0.1 $\pm$      0.0  &  28.3 $\pm$  20.4 &  0.67 $\pm$  0.44 &   0.2 $\pm$   0.1 &   0.2 $\pm$   0.1 &  67  & \ldots	 &  0        &   10.00 $\pm$   1.80 \\ 
73 &      6.1 &    -49.7 &    279.1 &    451.1 &      0.7 $\pm$      0.1 &     11.5 $\pm$      0.7  &      0.1 $\pm$      0.1  &  14.5 $\pm$  10.8 &  0.94 $\pm$  0.67 &   0.3 $\pm$   0.2 &   0.2 $\pm$   0.2 &  31  & \ldots	 &  3        &    0.05 $\pm$   0.00 \\ 
74 &      2.8 &    -35.9 &    355.8 &    432.6 &      0.4 $\pm$      0.1 &      6.2 $\pm$      0.6  &      0.1 $\pm$      0.1  &  20.8 $\pm$  15.8 &  0.74 $\pm$  0.52 &   0.2 $\pm$   0.2 &   0.2 $\pm$   0.1 &  37  & 0.8447	 &  5        &    4.35 $\pm$   1.98 \\ 
75 &      1.3 &     -9.4 &    502.5 &    424.4 &      3.2 $\pm$      0.1 &     30.8 $\pm$      1.1  &      1.7 $\pm$      0.1  &  88.5 $\pm$   8.1 & 15.44 $\pm$  1.10 &   5.1 $\pm$   0.4 &   4.0 $\pm$   1.2 & \ldots & \ldots &  \ldots & \ldots\\ 
76 &      0.0 &      0.2 &    556.2 &    417.2 &      4.6 $\pm$      0.1 &     76.1 $\pm$      1.5  &      0.9 $\pm$      0.2  &  19.7 $\pm$   3.7 &  8.49 $\pm$  1.50 &   2.8 $\pm$   0.5 &   2.2 $\pm$   0.8 &  45  & 0.8451	 &  0        &    1.41 $\pm$   0.11 \\ 
77 &     -3.2 &      1.5 &    563.1 &    399.2 &      1.7 $\pm$      0.1 &     13.6 $\pm$      0.9  &      1.0 $\pm$      0.1  & 121.9 $\pm$  17.1 &  9.37 $\pm$  0.90 &   3.1 $\pm$   0.3 &   2.4 $\pm$   0.8 & \ldots & \ldots &  \ldots & \ldots\\ 
78 &     -4.2 &    -57.9 &    233.5 &    393.7 &      7.0 $\pm$      0.2 &    154.6 $\pm$      1.7  &     -0.6 $\pm$      0.2  &  -6.0 $\pm$   1.9 & -5.23 $\pm$  1.68 &  -1.7 $\pm$   0.5 &  -1.4 $\pm$   0.6 & \ldots & \ldots &  \ldots & \ldots\\ 
79 &     -6.0 &     37.8 &    765.2 &    383.4 &      3.5 $\pm$      0.1 &     37.5 $\pm$      1.2  &      1.7 $\pm$      0.2  &  74.3 $\pm$   6.9 & 15.79 $\pm$  1.18 &   5.2 $\pm$   0.4 &   4.1 $\pm$   1.3 & \ldots & \ldots &  \ldots & \ldots\\ 
80 &     -7.5 &    -43.4 &    314.0 &    375.6 &      0.4 $\pm$      0.0 &      6.1 $\pm$      0.5  &      0.1 $\pm$      0.1  &  28.3 $\pm$  16.2 &  0.98 $\pm$  0.52 &   0.3 $\pm$   0.2 &   0.3 $\pm$   0.2 & \ldots & \ldots &  \ldots & \ldots\\ 
81 &     47.4 &    -47.1 &    293.4 &    680.4 &      0.4 $\pm$      0.0 &      6.0 $\pm$      0.5  &      0.1 $\pm$      0.1  &  17.5 $\pm$  14.7 &  0.59 $\pm$  0.47 &   0.2 $\pm$   0.2 &   0.2 $\pm$   0.1 & \ldots & \ldots &  \ldots & \ldots\\ 
82 &     -7.6 &    -81.0 &    105.0 &    375.0 &      0.5 $\pm$      0.1 &      7.0 $\pm$      0.6  &      0.2 $\pm$      0.1  &  36.0 $\pm$  16.0 &  1.43 $\pm$  0.57 &   0.5 $\pm$   0.2 &   0.4 $\pm$   0.2 & \ldots & \ldots &  \ldots & \ldots\\ 
83 &     70.7 &     -4.0 &    533.0 &    809.8 &      0.5 $\pm$      0.0 &      4.6 $\pm$      0.5  &      0.2 $\pm$      0.1  &  81.2 $\pm$  25.2 &  2.13 $\pm$  0.51 &   0.7 $\pm$   0.2 &   0.6 $\pm$   0.2 &  65  & \ldots	 &  1        &    5.03 $\pm$   1.53 \\ 
84 &     63.5 &     17.1 &    650.1 &    769.6 &      3.9 $\pm$      0.1 &     43.8 $\pm$      1.2  &      1.8 $\pm$      0.2  &  67.2 $\pm$   6.1 & 16.68 $\pm$  1.25 &   5.5 $\pm$   0.4 &   4.3 $\pm$   1.3 &  64  & \ldots	 &  10       &    0.07 $\pm$   0.01 \\ 
85 &     69.9 &    -12.2 &    487.0 &    805.2 &      0.4 $\pm$      0.0 &      6.1 $\pm$      0.5  &      0.1 $\pm$      0.1  &  30.6 $\pm$  16.3 &  1.06 $\pm$  0.52 &   0.3 $\pm$   0.2 &   0.3 $\pm$   0.2 & \ldots & \ldots &  \ldots & \ldots\\ 
86 &     42.4 &    -30.1 &    387.9 &    652.6 &      1.2 $\pm$      0.1 &     20.2 $\pm$      1.0  &      0.2 $\pm$      0.1  &  17.1 $\pm$   8.4 &  1.96 $\pm$  0.91 &   0.6 $\pm$   0.3 &   0.5 $\pm$   0.3 &  16  & \ldots	 &  5        &    0.88 $\pm$   0.02 \\ 
87 &     67.1 &     -5.9 &    522.2 &    789.6 &      0.9 $\pm$      0.1 &     21.2 $\pm$      0.8  &     -0.2 $\pm$      0.1  & -13.4 $\pm$   6.1 & -1.61 $\pm$  0.75 &  -0.5 $\pm$   0.2 &  -0.4 $\pm$   0.2 & \ldots & \ldots &  \ldots & \ldots\\ 
88 &     67.5 &    -56.9 &    239.1 &    791.8 &      1.0 $\pm$      0.1 &     23.5 $\pm$      0.9  &     -0.1 $\pm$      0.1  &  -7.3 $\pm$   6.5 & -0.97 $\pm$  0.89 &  -0.3 $\pm$   0.3 &  -0.3 $\pm$   0.2 & \ldots & \ldots &  \ldots & \ldots\\ 
89 &     62.1 &     39.4 &    773.8 &    762.1 &      3.6 $\pm$      0.1 &     48.4 $\pm$      1.3  &      1.2 $\pm$      0.1  &  41.0 $\pm$   5.2 & 11.23 $\pm$  1.25 &   3.7 $\pm$   0.4 &   2.9 $\pm$   0.9 & \ldots & \ldots &  \ldots & \ldots\\ 
90 &     51.6 &    -45.2 &    303.7 &    703.5 &     10.0 $\pm$      0.2 &    142.3 $\pm$      1.9  &      3.1 $\pm$      0.3  &  34.9 $\pm$   2.7 & 28.17 $\pm$  1.97 &   9.2 $\pm$   0.6 &   7.3 $\pm$   2.2 &  96  & \ldots	 &  10       &    3.15 $\pm$   1.75 \\ 
91 &     41.7 &     18.2 &    656.4 &    648.7 &      0.3 $\pm$      0.0 &      1.3 $\pm$      0.4  &      0.2 $\pm$      0.0  & 258.1 $\pm$ 112.1 &  1.88 $\pm$  0.38 &   0.6 $\pm$   0.1 &   0.5 $\pm$   0.2 & \ldots & \ldots &  \ldots & \ldots\\ 
92 &     54.2 &    -78.1 &    121.2 &    717.9 &      3.7 $\pm$      0.1 &     63.7 $\pm$      1.4  &      0.6 $\pm$      0.2  &  16.1 $\pm$   4.0 &  5.80 $\pm$  1.37 &   1.9 $\pm$   0.4 &   1.5 $\pm$   0.6 & \ldots & \ldots &  \ldots & \ldots\\ 
93 &     44.8 &    -23.6 &    423.6 &    665.9 &      0.8 $\pm$      0.1 &     14.3 $\pm$      0.8  &      0.1 $\pm$      0.1  &  16.8 $\pm$   9.6 &  1.36 $\pm$  0.74 &   0.4 $\pm$   0.2 &   0.4 $\pm$   0.2 & \ldots & \ldots &  \ldots & \ldots\\ 
94 &     49.6 &     35.1 &    750.3 &    692.5 &      2.3 $\pm$      0.1 &     41.7 $\pm$      1.2  &      0.3 $\pm$      0.1  &  11.2 $\pm$   5.0 &  2.65 $\pm$  1.14 &   0.9 $\pm$   0.4 &   0.7 $\pm$   0.4 & \ldots & \ldots &  \ldots & \ldots\\ 
\enddata 
\tablecomments{Columns:  (1) Name is CLJ0023+0423B$-$ followed by number listed in column.  (2) RA offset from BCG in arcseconds.  (3) DEC offset from BCG in arcseconds.  (4) Image x-position in pixels.  (5) Image y-position in pixels.  (6) Narrow-band flux in ADU/s. (7)  $J$-band flux in ADU/s. (8)  Continuum-subtracted flux in ADU/s.  (9) Narrow-band EW in \AA.  (10) Flux of \ha \ in units of 10$^{-17}$~ergs s$^{-1}$ cm$^{-2}$.  (11) Luminosity of \ha \ in units of 10$^{41}$~ergs s$^{-1}$.  (12) SFR in units of $h_{100}^{-2}\ \rm M_\odot yr^{-1}$.  (13) Catalog name used in \citeauthor{lubin98a} \citeyear{lubin98a}.  (14) Redshift from \citeauthor{postman98} \citeyear{postman98}.  (15) T-type from \citeauthor{lubin98a} \citeyear{lubin98a}, where E=0, S0=0.1, Sa=1, Sb=3, Sc=5, Sd=7, Sm/Im=9, Peculiar=10.  We denote galaxies that \citeauthor{lubin98a} classify as extremely compact with a T-type of $-$1 and
peculiar with 10.  (16) Sersic index.} 
\end{deluxetable} 

\end{turnpage}

\begin{deluxetable}{lccccccccccc}
\tabletypesize{\tiny}
\tablecaption{Integrated \ha \ SFRs of Galaxy Clusters \label{cljlitsfrs}}
\tablehead{\colhead{Name}  	&\colhead{z}  	& \colhead{$\sigma$} & \multicolumn{2}{c}{$R_{200}$} 
& \multicolumn{2}{c}{Survey Radius} & \colhead{Vol} & \colhead{$\Sigma$~SFR}& \colhead{$\Sigma$~SFR/M$_{cl}$} & \colhead{Tech} & \colhead{Ref.} \\
\colhead{(1)} & (2) & (3) & (4) & (5) & (6) & (7) & (8) & (9) & (10) & (11) &(12)}
\startdata
{CL J0023+0423B}&0.845  & 415$^{+102}_{-63}$  & \phn1.34  & 0.43 & 1.29 & 0.96 & 1.00 & 38.2$\pm$11.6    &58.4$\pm$17.7& I & 1\\
Abell 2390      & 0.228  &1023$\pm$200        & 10.29 	  & 1.58 & 8.00 & 0.78 & 1.00 & 47.0$\pm$16.5 	&\phn3.9$\pm$1.4& I & 2\\
AC 114 	& 0.32\phn       &1390$\pm$200 	      & \phn9.54  & 1.87 & 4.35 & 0.46 & 0.99 & \phn7.7$\pm$2.3 &\phn0.27$\pm$0.08& S & 3\\
Abell 1689  &0.183	 &1273$\pm$200 	      & 16.99  	  & 2.20 & 4.35 & 0.27 & 0.74 & 14.5$\pm$4.4 	&\phn0.84$\pm$0.25& S & 4\\
\enddata
\tablecomments{Columns: (1) Cluster name. 
(2) Redshift. 
(3) Velocity dispersion in $\rm km\ s^{-1}$. 
Velocity dispersions for AC~114 and Abell~1689 are calculated from $L_X$ using 
best-fit $L_X - \sigma$ relation of \citet{mahdavi01} because measured dispersions are inflated by substructure.
(4) $R_{200}$ in arcmin. 
(5) $R_{200}$ in \h \ Mpc.  
(6) Survey radius in arcmin. 
(7) Survey radius in units of $R_{200}$. 
(8) Fraction of volume imaged within $0.5 \times R_{200}$. 
(9) Integrated SFR in \smy. 
(10) Integrated SFR per cluster mass, in units of $h_{100}^{-3}\ \rm M_\odot \ yr^{-1}\ / \ 10^{14} \  M_\odot$.
(11) Observing Technique: I = narrow-band imaging, S = spectroscopic survey. 
(12) References.}
\tablerefs{(1) This work; (2) \citeauthor{balogh00} \citeyear{balogh00};
(3) \citeauthor{couch01} \citeyear{couch01}; (4) \citeauthor{balogh02} \citeyear{balogh02}.}
\end{deluxetable}

\end{document}